\documentclass{jfm}
\usepackage{graphicx}
\usepackage{epstopdf, epsfig}
\usepackage{amsmath}
\usepackage{bm}
\usepackage[dvipsnames]{xcolor}

\shorttitle{Fringing magnetic fields in thermal convection}
\shortauthor{S. Bhattacharya, T. Boeck, D. Krasnov and J. Schumacher}

\title{Effects of strong fringing magnetic fields on turbulent thermal convection}

\author{Shashwat Bhattacharya\aff{1}
  \corresp{\email{shashwat.bhattacharya@tu-ilmenau.de}},
  Thomas Boeck\aff{1},
  Dmitry Krasnov\aff{1}
 \and J{\"o}rg  Schumacher\aff{1}}

\affiliation{\aff{1}Institute of Thermodynamics and Fluid Mechanics, Technische Universit{\"a}t Ilmenau,
P.O. Box 100565, D-98684 Ilmenau, Germany}

\begin{document}

\maketitle

\begin{abstract}
We study the influence of fringing magnetic fields on turbulent thermal convection in a horizontally extended rectangular domain. The magnetic field is created in the gap between two semi-infinite planar magnetic poles, with the convection layer located near the edge of the gap. We employ direct numerical simulations in this setup for fixed Rayleigh and small Prandtl numbers, but vary the fringe-width by controlling the gap between the magnetic poles and the convection cell. The magnetic field generated by the magnets is strong enough to cease the flow in  high magnetic flux region of the convection cell. We observe that as the local vertical magnetic field strength increases, the large scale structures become thinner and align themselves perpendicular to the longitudinal sidewalls. We determine the local Nusselt and Reynolds numbers as functions of the local Hartmann number (based on the vertical component of the magnetic field) and estimate the global heat and momentum transport. We show that the global heat transport decreases with increasing fringe-width for strong magnetic fields \textcolor{black}{but increases} with increasing fringe-width for weak magnetic fields. In the regions of large vertical magnetic fields, the convective motion becomes confined to the vicinity of the sidewalls. The amplitudes of these wall modes show a non-monotonic dependence on the fringe-width. 
\end{abstract}

\section{Introduction}\label{sec:Introduction}
Flows that are driven by buoyancy forces are a common occurrence in nature as well as in technological applications. The driving mechanism is the temperature dependence of the fluid density, which leads to density variations when heat is transported through the fluid. A simplified paradigm for such flows is the Rayleigh-B{\'e}nard convection (RBC), which consists of a fluid layer that is heated from below and cooled from above~\citep{Chandrasekhar:book:Instability}. While the understanding of RBC has increased substantially in the past decades \citep[see, for example,][]{Ahlers:RMP2009,Chilla:EPJE2012,Verma:BDF}, it must be noted that a  multitude of additional forces, such as those generated by rotation and magnetic fields, can affect buoyancy-driven convection in nature or in industrial applications. The effects of these forces have been relatively less explored. The present study deals with magnetoconvection, that is, thermal convection of electrically conducting fluids under the effect of magnetic fields.

In magnetoconvection, flows are acted upon by buoyancy as well as by Lorentz forces generated due to a magnetic field~\citep{Weiss:book}. Convection in the Sun and planetary dynamos are examples of magnetoconvection occurring in nature where the magnetic field is maintained by the flow. In technological applications, the magnetic field is usually not primarily caused by the flow but externally imposed. Examples are liquid metal batteries for renewable energy storage, growth of semiconductor monocrystals, flow control of hot metal melts by electromagnetic brakes in metallurgy, and heat transfer in blankets in nuclear fusion reactors. 

Magnetoconvection is governed by the equations for conservation  of mass,  momentum and  energy as well as Maxwell's equations and Ohm's law. The governing nondimensional parameters of magnetoconvection are i) Rayleigh number $\Ray$ -- the ratio of buoyancy to dissipative forces, ii) Prandtl number $\Pran$ -- the ratio of kinematic viscosity to thermal diffusivity, iii) Hartmann number $\Ha$ -- the ratio of Lorentz to viscous force, and iv) the magnetic Prandtl number $\Pm$ -- the ratio of kinematic viscosity to magnetic diffusivity. \textcolor{black}{Instead of Hartmann number, the free-fall interaction parameter $N_f = \Ha^2 \sqrt{\Pran/\Ray}$, also called the Stuart number, can be used as a governing parameter~\citep{Liu:JFM2018,Zuerner:JFM2020,Xu:JFM2022}. The free-fall interaction parameter represents the ratio of magnetic and buoyant forcing in the fluid.} The important nondimensional output parameters of magnetoconvection are i) the Nusselt number $\Nu$, which quantifies the global heat transport, ii) the Reynolds number $\Rey$ -- the ratio of inertial forces to viscous forces, and iii) the magnetic Reynolds number $\Rm$ -- the  ratio of induction to diffusion of the magnetic field. For sufficiently small magnetic Reynolds numbers, the induced magnetic field is negligible compared to the applied magnetic field and is therefore neglected in the expressions of Lorentz force and Ohm's law~\citep{Roberts:book,Davidson:book:MHD,Verma:ROPP2017,Verma:ET}. In such cases, referred to as quasi-static magnetoconvection, the induced magnetic field adjusts instantaneously to the changes in velocity. In the {\em quasi-static approximation}, there exists a one-way influence of the magnetic field on the flow only.

Magnetoconvection has been studied theoretically in the past~\citep[for example,][]{Chandrasekhar:book:Instability,Houchens:JFM2002,Busse:PF2008} as well as with the help of experiments~\citep[for example,][]{Nakagawa:PRSL1957,Fauve:JPL1981,Cioni:PRL2000,Aurnou:JFM2001,Burr:PF2001,King:PNAS2015,Vogt:PRF2018,Vogt:JFM2021,Zuerner:JFM2020,Grannan:JFM2022} and numerical simulations~\citep[for example,][]{Liu:JFM2018,Yan:JFM2019,Akhmedagaev:MHD2020,Akhmedagaev:JFM2020,Nicoski:PRF2022}. A horizontal magnetic field has been found to cause the large-scale rolls to become quasi two-dimensional and align in the direction of the field~\citep{Fauve:JPL1981,Busse:JTAM1983,Burr:JFM2002,Yanagisawa:PRE2013,Tasaka:PRE2016,Vogt:PRF2018,Vogt:JFM2021}. These self-organized flow structures reach an optimal state characterised by a significant increase in heat transport and convective velocities~\citep{Vogt:JFM2021}. In contrast, strong vertical magnetic fields suppress convection~\citep{Chandrasekhar:book:Instability,Cioni:PRL2000,Zuerner:JFM2020,Akhmedagaev:MHD2020,Akhmedagaev:JFM2020} with the flow ceasing above a critical Hartmann number. At this threshold, the flow in the center is fully suppressed with convective motion only in the vicinity of sidewalls~\citep{Busse:PF2008,Liu:JFM2018,Zuerner:JFM2020,Akhmedagaev:MHD2020,Akhmedagaev:JFM2020} if existent. These so-called wall modes, which are also present in confined rotating convection~\citep{Zhong:PRL1991,Ecke:EPL1992,Goldstein:JFM1993,Goldstein:JFM1994,Liu:PRE1999,King:JFM2012}, are particularly relevant in technical applications in closed vessels. 
\citet{HurlBurt:APJ1996} and  \citet{Nicoski:PRF2022} studied convection with tilted magnetic fields. The results of \citet{HurlBurt:APJ1996} suggest that the mean flows tend to travel in the direction of the tilt. \citet{Nicoski:PRF2022}  reported qualitative similarities between convection with tilted magnetic field and that with vertical magnetic field in terms of the behavior of convection patterns, heat transport, and flow speed. 

\textcolor{black}{It must be noted that all the aforementioned works on magnetoconvection have been restricted to uniformly imposed magnetic field, which is an idealized approximation. However, in engineering applications such liquid metal batteries~\citep{Kelley:AMR2018}, cooling blankets in fusion reactors~\citep{Mistralengo:FED2021}, electromagnetic stirring~\citep{Davidson:ARFM1999}, electromagnetic brakes~\citep{Davidson:book:MHD}, and non-contact flow measurements involving Lorentz force velocimetry~\citep{Thess:PRL2006}, the imposed magnetic fields are localized and thus vary in space. 
Further, strong homogeneous fields in large regions of space can only be generated by magnets of large size which are difficult to design and very costly to build and operate~\citep{Barleon:KIT1996}. Thus, it is important to understand the impact of spatially varying magnetic fields on magnetohydrodynamic flows. Although there are several studies on the channel flow of liquid metals under the influence of nonhomogeneous magnetic fields~\citep[see, for example,][]{Sterl:JFM1990,Votyakov:PF2009,Moreau:PMCPB2010,Albets-Chico:FED2013,Klueber:FED2020}, no similar work on magnetoconvection has been conducted so far to the best of our knowledge.}

Convection flows in horizontally extended domains are often organized into prominent and coherent large-scale patterns that persist for very long times and can extend over scales in the lateral direction that are much larger than the domain height. These so-called superstructures have a strong influence on the turbulent transport properties of the flow. A prominent example of such superstructures is the granulation at the surface of the Sun \citep{Schumacher:RMP2020}. While the understanding of the process of formation and evolution of these superstructures and their characteristic scales in the absence of magnetic fields has improved significantly over the recent years \citep[see, for example,][]{Pandey:NC2018,Stevens:PRF2018,Krug:JFM2020}, the case with inhomogeneous magnetic fields still leaves several questions open as the physical complexity is increased.

In the present work, we attempt to fill some of these gaps and study the effect of strong spatially varying magnetic fields in turbulent convection under the quasi-static approximation. The effects of localized magnetic fields on turbulent superstructures and their impact on the local and global turbulent heat and momentum transport are explored. We consider a Rayleigh-B{\'e}nard cell with fringing magnetic fields generated by semi-infinite magnetic poles. \textcolor{black}{The strength of the magnets is such that convection in the regions of strong magnetic fields is fully suppressed in the bulk. Thus, in a single convection cell, we obtain different local regimes of magnetoconvection, ranging from the turbulent regime in the regions of weak magnetic flux to wall-attached convection in the regions of strong magnetic flux region. As mentioned earlier, the superstructures are formed in horizontally-extended domains; hence, the convection cell employed in our present study has a large aspect ratio.} We study the spatial variation of size and orientation of turbulent superstructures along with the turbulent transport inside the convection cell. We also study wall-attached convection in the regions of strong magnetic flux regions, the wall modes. Although the setup is a simple way to study the influence of spatially varying magnetic fields on convection, it permits us to carry out such a parametric study systematically. This is the main motivation of the present work. Future studies can be conducted in a similar manner for more complex arrangements.

The outline of the paper is as follows. Section~\ref{sec:Model} describes the magnetoconvection setup along with the governing equations of magnetoconvection and the spatial distribution of the magnetic field. The details of our numerical simulations are also discussed in \S~\ref{sec:Model}. In \S~\ref{sec:Results}, we present the numerical results, detailing the behavior of large-scale structures, the spatial distribution of heat and momentum transport along with their variations with the magnetic field, the dependence of the global heat and momentum transport on the fringe-width, and wall-attached convection. We conclude in \S~\ref{sec:Conclusion} with a summary and an outlook.

\section{Numerical model}\label{sec:Model}
\subsection{Problem setup and equations}\label{sec:Problem}
 
\begin{figure}
  \centerline{\includegraphics[scale = 0.3]{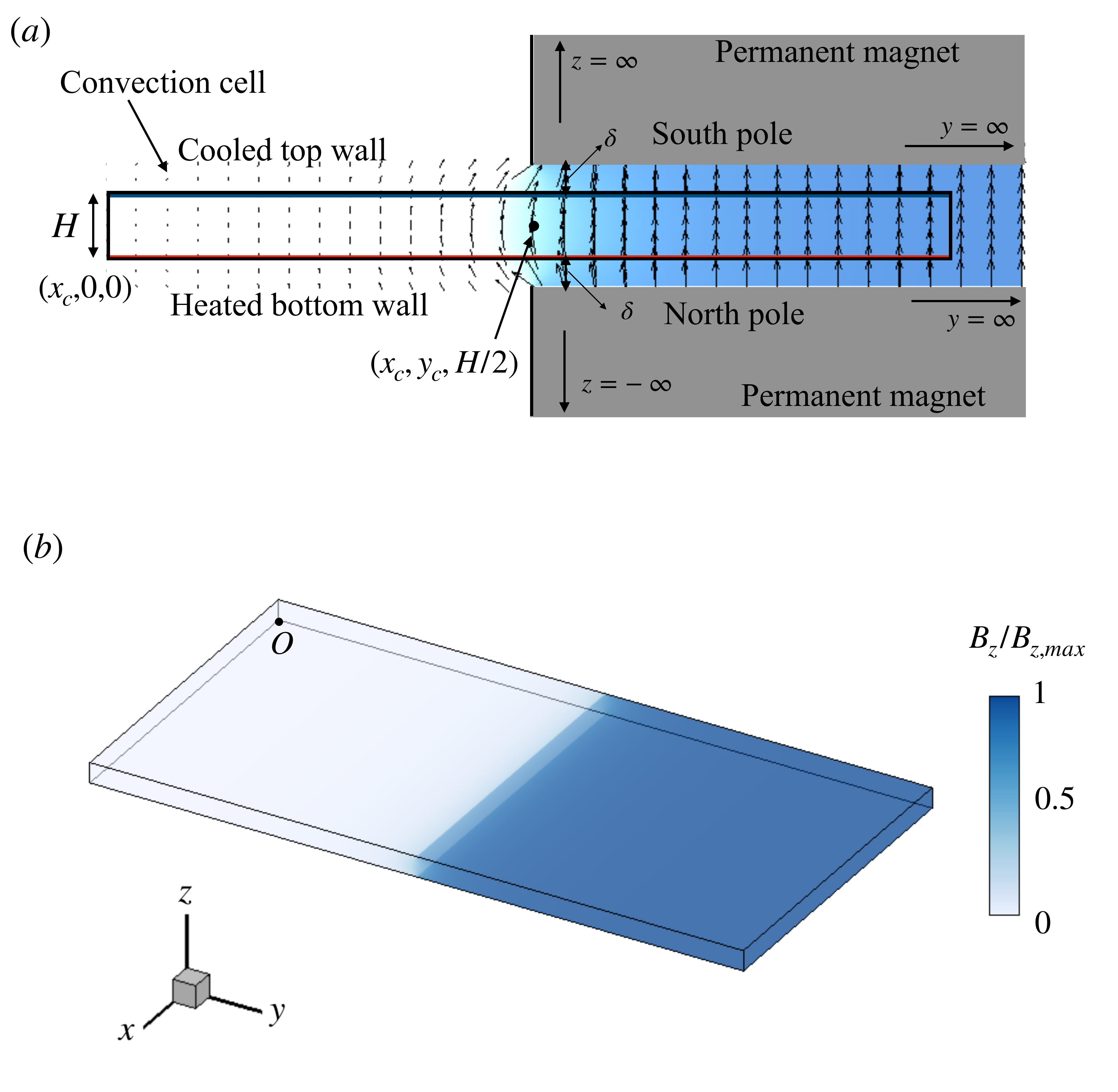}}
  \caption{\textcolor{black}{(a) Schematic of the vertical section at $x=l_x/2$ of the proposed magnetoconvection arrangement. The semi-infinite permanent magnets (colored in grey) generate the localized magnetic field represented by the vector plots drawn in the convection cell. The magnetic field distribution is described by Eqs.~(\ref{eq:Bx}) to (\ref{eq:Bz}). (b) A three-dimensional view of the convection cell used in our study with the density contours of the vertical component of the normalized imposed magnetic field for $\delta/H=0.01$. Here, $O$ refers to the point of the origin, and $\delta$ is the gap between the magnetic pole and the convection cell.}}
\label{Fig:Magnetoconvection_setup}
\end{figure}

We consider a horizontally extended convection cell of size $l_x \times l_y \times H$ which is under the influence of a magnetic field generated by two semi-infinite permanent magnets.  The north pole of one magnet faces the bottom of the convection cell and the south pole of the second magnet faces the top of the convection cell. These magnets extend from $-\infty$ to $\infty$ in the $x$-direction, $l_y/2$ to $\infty$ in the $y$-direction, from near the top wall to $\infty$ in the positive $z$-direction, and from near the bottom wall to $- \infty$ in the negative $z$ direction. A schematic of \textcolor{black}{a vertical section of} the arrangement is provided in figure~\ref{Fig:Magnetoconvection_setup}(a).  Using the model of \citet{Votyakov:TCFD2009} for permanent magnets, we obtain the following relations for the non-dimensional imposed magnetic field $\boldsymbol{B}_0=\{B_x, B_y, B_z \}$ generated in the above configuration:
\begin{eqnarray}
B_x &=& 0, \label{eq:Bx}\\
B_y &=& -{\color{black}\frac{1}{4\pi}\Bzmax}\ln \left[ \frac{(y - y_c)^2 - \{(z - z_c) - (H/2 + \delta)\}^2}{(y - y_c)^2 + \{(z - z_c) + (H/2+\delta)\}^2} \right], \label{eq:By}\\
B_z &=& -{\color{black}\frac{1}{2\pi}\Bzmax} \arctan \left [\frac{2(H/2 + \delta)(y-y_c)}{(y-y_c)^2 + (z-z_c)^2 - (H/2+\delta)^2} \right ], \label{eq:Bz}
\end{eqnarray}
where $y_c$ and $z_c$ are the $y$ and $z$ coordinates, respectively, of the cell-center, \textcolor{black}{$\Bzmax$ is the maximum limiting value of the vertical component of the magnetic field at $y \rightarrow \infty$}, and $\delta$ is the gap between the permanent magnet and the top/bottom wall (see figure~\ref{Fig:Magnetoconvection_setup}). The aforementioned magnetic field distribution satisfies ${\bm\nabla}\times {\bm B}_0=0$ and ${\bm\nabla}\cdot {\bm B}_0=0$.  In the above configuration, the magnetic field is almost absent in nearly half of the RBC cell (for $y \lesssim l_y/2$), increases steeply about the midplane ($y \approx l_y/2$), and then becomes strong in the other half of the cell (for $y \gtrsim l_y/2$). 
\textcolor{black}{A three-dimensional view of the convection cell used in our study with the density contours of the vertical component of the imposed magnetic field is shown in figure~\ref{Fig:Magnetoconvection_setup}(b).}
The region where the gradient of the magnetic field is large is called the fringe zone; \textcolor{black}{we provide a mathematical definition of the fringe zone later in this subsection.} 
The region of the convection cell that is outside the gap between the magnets will be henceforth referred to as {\em weak magnetic flux region} and the one inside the gap will be referred to as {\em strong magnetic flux region}. 

\begin{figure}
  \centerline{\includegraphics[scale = 0.35]{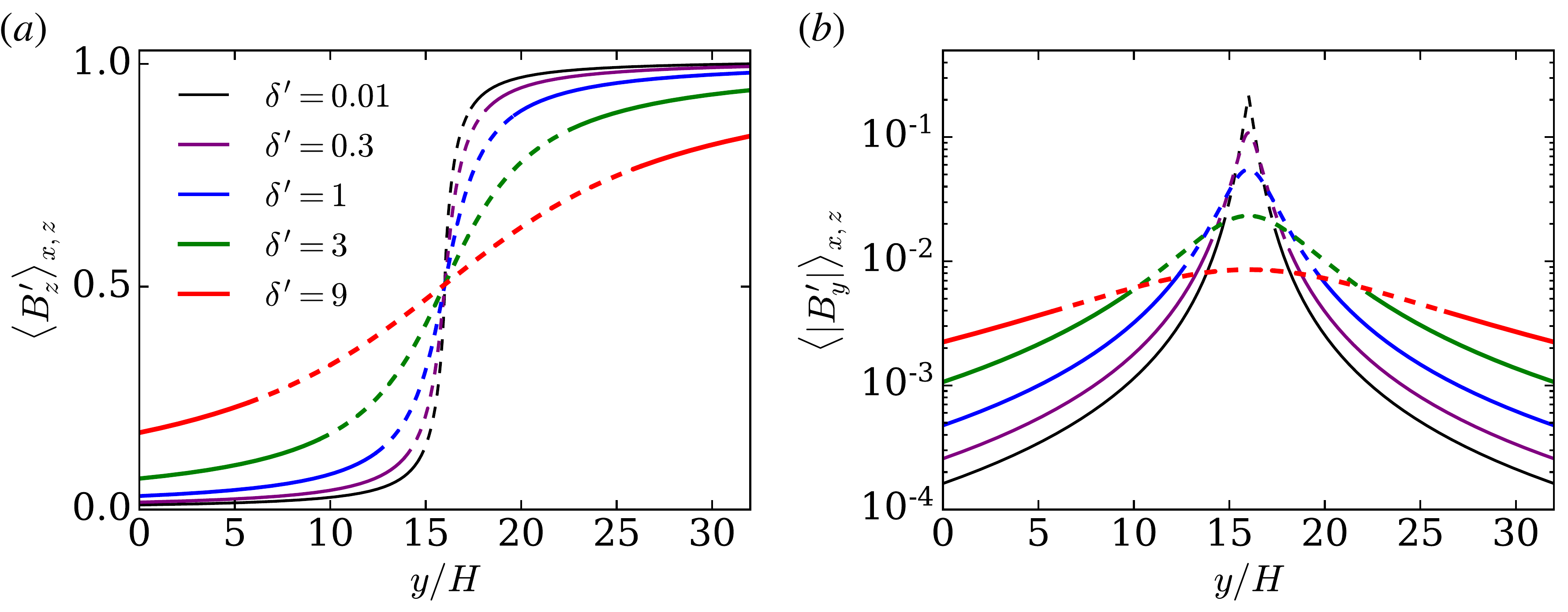}}
  \caption{\textcolor{black}{Distribution of the profiles of  (a) vertical magnetic field and (b) absolute value of the horizontal magnetic field, both averaged over $x$ and $z$, along the lengthwise direction (normalized by the height $H$ of the cell) for different values of the normalized gap ($\delta' = \delta/H$) between the magnetic poles and the thermal plates. The magnetic fields are normalized by the maximum value of the vertical magnetic field ($B_z'=B_z/\Bzmax$ and $B_y'=B_y/\Bzmax$). The curves are represented as dashed lines in the fringe zone and as solid lines outside the fringe zone.}} 
\label{fig:Bz_By_y}
\end{figure}

Equations (\ref{eq:By}) and (\ref{eq:Bz}) imply that the spatial distribution of the magnetic field is strongly dependent on the gap width ($\delta$) between the magnetic poles and the convection cell. 
\textcolor{black}{In figures~\ref{fig:Bz_By_y}(a) and (b), we plot, respectively, the profiles of the normalized mean vertical magnetic field ($\langle B_z/\Bzmax \rangle_{x,z}$) and the normalized mean horizontal magnetic field ($\langle |B_y|/\Bzmax \rangle_{x,z}$) versus the normalized lengthwise coordinate $y/H$. In the above, $|\cdot|$ and $\langle \cdot \rangle_{x,z}$, respectively, represent the absolute value and averaging over $x$ and $z$.} As evident in figure~\ref{fig:Bz_By_y}(a), the gradient of the vertical magnetic field decreases as $\delta$ is increased.
\textcolor{black}{In the present setup, we define the \emph{fringe zone} as the region in which $0.1 < \widehat{B}_z < 0.9$, where
\begin{equation}
 \widehat{B}_z = \frac{\langle B_z \rangle_{x,z} - \Bzmind}{\Bzmaxd - \Bzmind}.   
\end{equation}
In the above, $\Bzmaxd$ and $\Bzmind$ are, respectively, the maximum and the minimum values of the vertical magnetic field in the convection cell \footnote{\textcolor{black}{It must be noted that $\Bzmaxd \neq\Bzmax$; $\Bzmaxd$ is the maximum value of $B_z$ inside the convection cell, and $\Bzmax$ is the maximum limiting value of $B_z$ at $y \rightarrow \infty$.}}. In figures~\ref{fig:Bz_By_y}(a) and (b), the curves describing the magnetic field profiles are represented as dashed lines in the fringe zone and as solid lines outside the fringe zone. The figures clearly show that}
the width of the fringe zone increases with an increase in $\delta$. Further, the horizontal component of the magnetic field on the lateral vertical midplane at $y=l_y/2$ increases with a decrease in $\delta$. These variations in the magnetic field profile affect the convection patterns along with the associated global heat transport and will be studied in detail in the later sections.

The study will be conducted under the quasi-static approximation, in which the induced magnetic field is neglected as it is very small compared to the applied magnetic field. This approximation is fairly accurate for magnetoconvection in liquid metals~\citep{Davidson:book:MHD}. 
The nondimensionalized governing equations are as follows:
\begin{eqnarray}
\nabla \cdot \boldsymbol{u}&=&0 \label{eq:continuity} \\
\frac{\partial \boldsymbol{u}}{\partial t}  + \boldsymbol{u}\cdot \nabla \boldsymbol{u} &=& -\nabla p + T\hat{z}+ \sqrt{\frac{\Pran}{Ra}} \nabla^2 \boldsymbol{u}+ \Hazmax^2\sqrt{\frac{\Pran}{Ra}}(\boldsymbol{j} \times \tilde{\boldsymbol{B}}),
\label{eq:Momentum} \\
\frac{\partial T}{\partial t} + \boldsymbol{u} \cdot \nabla T &=& \frac{1}{\sqrt{Ra \Pran}}\nabla^2 T, \label{eq:T_energy} \\
\boldsymbol{j} &=& -\nabla \phi + (\boldsymbol{u} \times \tilde{\boldsymbol{B}}),
\label{eq:Current}\\
\nabla^2 \phi &=& \nabla \cdot (\boldsymbol{u} \times \tilde{\boldsymbol{B}}), 
\label{eq:Potential}
\end{eqnarray}
where $\boldsymbol{u}$, $\boldsymbol{j}$, $p$, $T$, and $\phi$ are the fields of velocity, current density, pressure, temperature, and electrical potential respectively, and  $\tilde{\boldsymbol{B}}$ is the applied magnetic field normalized by $\Bzmax$. The governing equations are made dimensionless by using the cell height $H$, the imposed temperature difference $\Delta$, and the free-fall velocity $U = \sqrt{\alpha g \Delta H}$ (where $g$ and $\alpha$ are, respectively, the gravitational acceleration and the volumetric coefficient of thermal expansion of the fluid). \textcolor{black}{The nondimensional governing parameters are the Rayleigh number ($\Ray$), the Prandtl number ($\Pran$), the Hartmann number ($\Hazmax$) based on $\Bzmax$, and the normalized gap ($\delta'$) between the thermal plates and the magnetic poles.} These parameters are defined as follows:

\begin{equation}
\Ray = \frac{\alpha g \Delta H^3}{\nu \kappa}, \quad \Pran= \frac{\nu}{\kappa}, \quad \Hazmax = \Bzmax H\sqrt{\frac{\sigma}{\rho \nu}}, \quad \textcolor{black}{\delta' = \frac{\delta}{H}},
\end{equation}
where $\nu$ is the kinematic viscosity, $\kappa$ is the thermal diffusivity, $\rho$ is the density, and $\sigma$ is the electrical conductivity of the fluid.
\textcolor{black}{For the sake of brevity, we henceforth omit the prime from $\delta’$.}
In the next section, we describe the simulation details.

\subsection{Numerical method}\label{sec:Numerical_method}
We conduct direct numerical simulations of \textcolor{black}{our magnetoconvection setup described in \S~\ref{sec:Problem}}.
 The spatial distribution of the magnetic field is given by equations (\ref{eq:Bx}) to (\ref{eq:Bz}).
We use a second-order finite difference code developed by \citet{Krasnov:CF2011} to numerically solve equations (\ref{eq:continuity}) to (\ref{eq:Potential}). 
\textcolor{black}{The Prandtl number $\Pran$ is chosen to be 0.021, which is the same as that of mercury. The Rayleigh number is fixed at $\Ray=10^5$. 
For our chosen $\Pran$, the aforementioned value of $\Ray$ causes significant turbulence in the regions of weak magnetic flux.
For $\Ray=10^5$, the critical Hartmann number (above which the bulk convection is suppressed) computed using the linear stability analysis of \citet{Chandrasekhar:book:Instability} for uniform vertical magnetic field is
\begin{equation}
\Ha_{z,c}=88.7.
\label{eq:Hac}
\end{equation}
In order to include the wall-attached convection regime in our analysis, we choose the maximum Hartmann number to be $\Hazmax = 120$, which is slightly higher than the critical Hartmann number.}

\textcolor{black}{As mentioned in \S~\ref{sec:Introduction}, we intend to study the behavior of turbulent superstructures in our present study. For our chosen values of $\Ray$ and $\Pran$, the typical length scales of the superstructures are 3 to 4 times the height of the domain~\citep{Pandey:NC2018}. Therefore, in order to obtain good statistics, we choose the horizontal extent of our domain to be around 5 times the expected length-scale of the superstructures. Furthermore, in order to capture the variation of the superstructures and the global heat transport with magnetic field strength more effectively, it is preferred that the domain is elongated along the $y$-direction, that is, along the gradient of the magnetic field. Therefore, we choose the domain-size to be $l_x \times l_y \times H = 16 \times 32 \times 1$. }

\textcolor{black}{We employ a grid-resolution of $4800 \times 9600 \times 300$ points.}
The mesh is non-uniform in the $z$-direction with stronger clustering of the grid points near the top and bottom boundaries. The elliptic equations for pressure, electric potential, and the temperature are solved based on applying cosine transforms in $x$ and $y$-directions and using a tridiagonal solver in the $z$-direction. The diffusive term in the temperature transport equation is treated implicitly. The time discretization of the momentum equation uses the fully explicit Adams-Bashforth/Backward-Differentiation method of second
order \citep{Peyret:book}. A constant time step size of $1 \times 10^{-4}$ free fall time units was chosen for our simulations, which satisfied the Courant–Friedrichs–Lewy (CFL) condition for all runs. 

All the walls are rigid and electrically insulated such that the electric current density $\boldsymbol{j}$ forms closed field lines inside the cell. The top and bottom walls are held fixed at $T=-0.5$ and $T=0.5$ respectively, and the sidewalls are adiabatic with $\partial T/\partial \eta =0$ (where $\eta$ is the component normal to sidewall). \textcolor{black}{All the simulations are initialized with the linear conduction profile for temperature (which is a function of the $z$-coordinate only) and a random noise of amplitude $A=0.001$ along the $z$-direction for velocity. We run the simulations initially on a coarse grid of $480 \times 960 \times 30$ points for 100 free-fall time units in which they converge to a statistically steady state. Following this, we successively refine the mesh to the required resolution of $4800 \times 9600 \times 300$ grid points and  allow the simulations to converge after each refinement. Once the simulations reach the statistically steady state at the highest resolution, they are run for a further 20--21 free-fall time units and a snapshot of the flow field is saved after every free-fall time unit.} 
\begin{table}
  \begin{center}
\def~{\hphantom{0}}
  \begin{tabular}{lcccc}
      Runs  & $\delta$   &   $\Rey_{global}$ & $\Nu_{global}$ & $t_N$\\[3pt]
       1   & 0.01 & $763 \pm 2$ & $1.80 \pm 0.01$ & 20\\
       2   & 0.3 & $767 \pm 5$ & $1.82 \pm 0.01$ & 20\\
       3  & 1 & $715 \pm 4$ & $1.75 \pm 0.01$ & 21\\
       4   & 3 & $644 \pm 3$ & $1.72 \pm 0.01$ & 20\\
       5 & 9 & $474 \pm 3$ & $1.60 \pm 0.01$ & 20\\
  \end{tabular}
  \caption{Parameters of the simulations: the gap ($\delta$) between the magnetic poles and the conducting walls, the  global Reynolds number ($\Reglobal$), the  global Nusselt number ($\Nuglobal$), and the number of free-fall time units ($t_N$) run by the solver after attaining statistically steady state. $\Reglobal$ and $\Nuglobal$ are averaged over $t_N$ timeframes and the errors are the standard deviations of the above quantities. The Rayleigh number, Prandtl number, and the maximum Hartmann number are fixed to $\Ray=10^5$, $\Pran=0.021$, and $\Ha_{z,max}=120$ respectively.}
  \label{table:Simulation}
  \end{center}
\end{table}

Table~\ref{table:Simulation} lists the important parameters of our simulation runs. In this table, we also report the turbulent momentum transfer quantified by the global Reynolds number ($\Rey_{global}$) and the turbulent heat transport quantified by the global Nusselt number ($\Nu_{global}$). These quantities are given by
\begin{eqnarray}
\Rey_{global} &=& u_{rms}\sqrt{\Ray / \Pran}, \label{eq:ReDirect} \\
\Nu_{global} &=& 1 + \sqrt{\Ray \Pran}\langle u_z T\rangle_{V,t_N}, \label{eq:NuDirect}
\end{eqnarray}
where $\langle \cdot \rangle$ represents averaging, $u_{rms} = \sqrt{\langle u_x^2 + u_y^2 + u_z^2 \rangle_{V,t_N}}$ with $V$ being the volume of the convection cell, and $t_N$ (also reported in table ~\ref{table:Simulation}) is the number of free-fall times run by the solver after attaining a statistically steady state. 

\begin{figure}
  \centerline{\includegraphics[scale = 0.4]{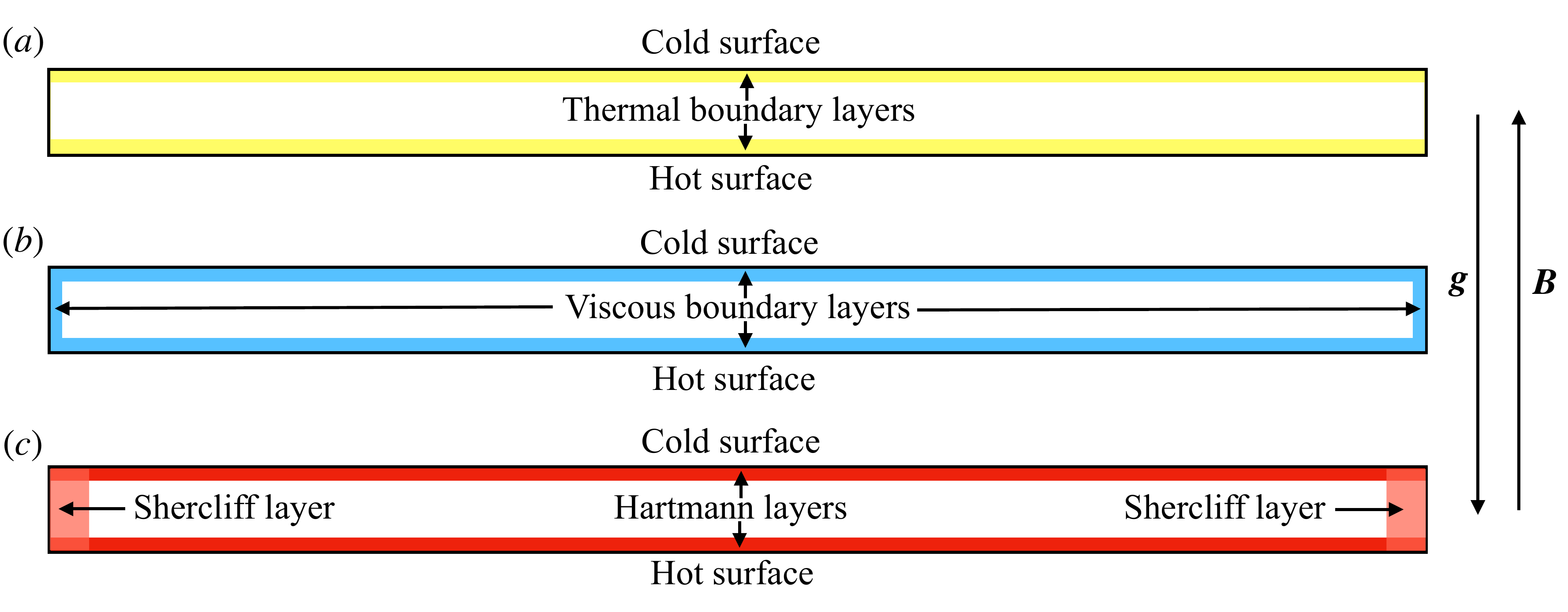}}
  \caption{Sketches of the vertical \textcolor{black}{($x$-$z$)} cross section of a Rayleigh-B{\'e}nard convection cell \textcolor{black}{(with aspect ratio similar to ours)} under vertical magnetic field showing (a) thermal boundary layers, (b) viscous boundary layers, and (c) Hartmann and Shercliff layers. } 
\label{fig:Boundary_layers}
\end{figure}

In turbulent magnetoconvection with vertical magnetic fields, four boundary layers are formed: 1) viscous boundary layers near all the walls, 2) thermal boundary and 3) Hartmann layers near the top and bottom walls, and 4) Shercliff layers near the sidewalls. Figure~\ref{fig:Boundary_layers} exhibits a sketch of the different types of boundary layers. In order to obtain accurate results, it is important that all these layers are adequately resolved. In our cases, although the imposed magnetic field has a horizontal component as well, it is very small compared to the vertical component near the sidewalls. Hence,  the sidewalls have been considered to contain only the Shercliff layers and no Hartmann layers. It must also be noted that the thicknesses of all the boundary layers vary along the $y$-direction because they depend on the local Hartmann number, which, in turn, affects the Reynolds and Nusselt numbers. Thus, for a conservative resolution analysis, the minimum thickness of these boundary layers has been considered. These are given by
\begin{align}
\delta_{T,min} &= \frac{1}{2\Nu_{max}}\,,\\ 
\delta_{u,min} &= \frac{1}{4\sqrt{\Rey_{max}}}\,,\\
\delta_{H,min} &= \frac{1}{\Hazmax}\,,\\  
\delta_{S,min} &= \frac{1}{\sqrt{\Hazmax}}\,,
\end{align}
where $\delta_{T,min}$, $\delta_{u,min}$, $\delta_{H,min}$, and $\delta_{S,min}$ are the minimum thicknesses of thermal boundary layers, viscous boundary layers, Hartmann layers, and the Shercliff layers, respectively. Here, we also remark that although the relation for the viscous boundary layer thickness given by $\delta_u \sim \Rey^{-1/2}$ is not very accurate \citep[see, for example,][]{Breuer:PRE2004,Scheel:JFM2012,Bhattacharya:PF2021}, it provides a reasonable estimate for small Prandtl number convection. \textcolor{black}{Further, $\Numax$ and $\Remax$, respectively, denote the Nusselt and Reynolds numbers in the regions where the magnetic fields are weak. They are computed as follows. For all our simulation cases, we identify the regions where $\widehat{B}_z < 0.1$ and compute the time-averaged Nusselt and Reynolds number integrated over these regions for every case. The maximum values of the above Nusselt and Reynolds  numbers (which turned out to be for the case with $\delta=0.3$) are taken to be $\Numax$ and $\Remax$, respectively, and are observed to be $\Numax=2.66$ and $\Remax=1115$.}  Based on the above values, we have a minimum of 12 points, 76 points, and 12 points, respectively, in the viscous boundary layers, thermal boundary layers, and in the Hartmann layers adjacent to the top and bottom walls; and 28 points in the Shercliff layer adjacent to the sidewalls. Thus, our simulations are well resolved and also satisfy the resolution criterion of \citet{Grotzbach:JCP1983} and \citet{Verzicco:JFM2003}.   
In the next section, we discuss our results in detail.
 
\section{Results}\label{sec:Results}
In this section, we discuss the characteristics of the large-scale convection patterns, the spatial profile of the large-scale momentum and the global heat transport, and the wall-attached convection in the regions of strong magnetic flux. For our analysis, we introduce the local vertical Hartmann number $\Ha_z(y)$ which is defined as,
\begin{equation}
    \Ha_z(y) = \Hazmax \frac{\langle B_z \rangle_{x,z}}{\Bzmax},
    \label{eq:localHa}
\end{equation}
where $B_z$ is the vertical component of the local magnetic field.
Thus, $\Ha_z(y)$ quantifies the strength of $B_z$ averaged over the corresponding $x$-$z$-plane. We will discuss the variations of the heat and momentum transport with $\Ha_z(y)$ and their resulting impact on the global dynamics.

\subsection{Large-scale structures} \label{sec:Superstructures}
\begin{figure}
  \centerline{\includegraphics[scale = 0.35]{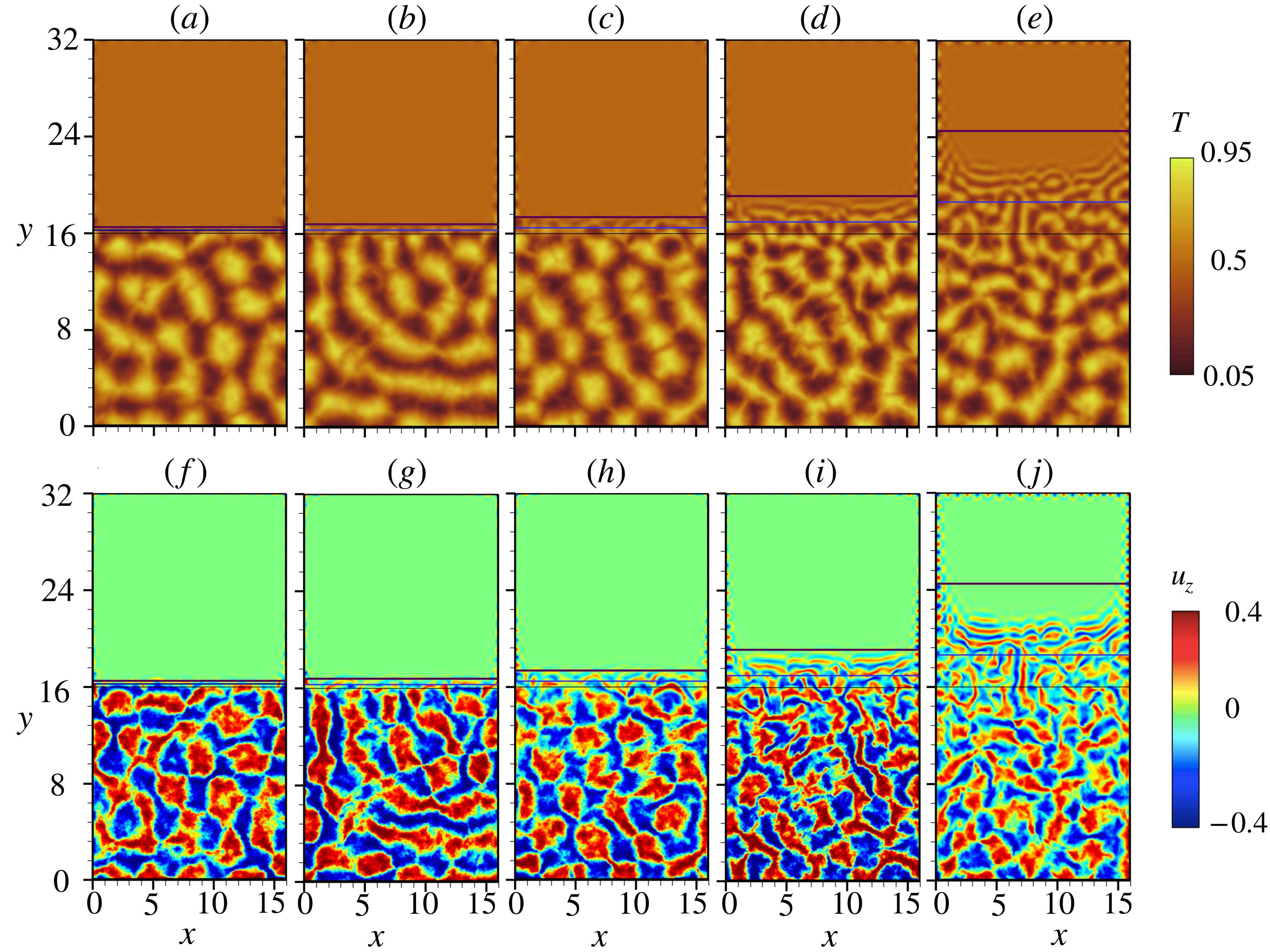}}
  \caption{Contour plots of time-averaged fields on the horizontal midplane for different fringe-widths of the imposed magnetic field. Plots of the temperature field for (a) $\delta=0.01$, (b) $\delta=0.3$, (c) $\delta=1$, (d) $\delta=3$, and (e) $\delta=9$. Plots of the vertical velocity field for (f) $\delta=0.01$, (g) $\delta=0.3$, (h) $\delta=1$, (i) $\delta=3$, and (j) $\delta=9$. The flows are organized into large-scale patterns that extend over scales larger than the domain height. \textcolor{black}{The positions corresponding to $\Ha_z(y)=\Ha_{z,c}$, $\Ha_z(y)=0.8 \Ha_{z,c}$, and the edge of the magnetic poles are represented by purple, blue, and black horizontal lines, respectively, in decreasing order of thickness. The magnetic poles extend from $y=16$ to $y=\infty$.}}
\label{fig:Superstructures}
\end{figure}

We use our numerical data to analyse the flow structures in the convection cell for different fringe-widths. Figures~\ref{fig:Superstructures}(a)-(e) exhibit the \textcolor{black}{contour} plots of temperature field on the horizontal midplane for $\delta=0.01$ to 9. The corresponding \textcolor{black}{contour} plots of the vertical velocity field are shown in figures~\ref{fig:Superstructures}(f)-(g). The data for the above plots are averaged over 20 to 21 timeframes (see table~\ref{table:Simulation}). The figures show that bulk convection is completely suppressed in the regions of strong magnetic fields.  
The region of suppressed bulk convection becomes smaller as $\delta$ is increased. This is expected because for large $\delta$, the gradient of the magnetic field is small and hence the region with $\Ha_z(y)$ above the critical Hartmann number is also small. 

Figures~\ref{fig:Superstructures}(a)--(j) also show that in the regions of $\Ha_z(y)<\Hazmax$, the flow gets organized into superstructures. 
The superstructures observed in the above figures in the weak magnetic flux regions are qualitatively similar to those observed by \cite{Pandey:NC2018} for $\Pran=0.021$. 
Further, the superstructures are relatively isotropic in the regions of weak magnetic flux, that is, they do not show any preference towards specific orientations. However, as the strength of the vertical magnetic field increases, the superstructures become more elongated and align themselves perpendicular to the sidewalls. The change in spatial structure of the flow is more clearly visible for $\delta=9$, in which the vertical magnetic field changes gradually. The transition in the orientation of the superstructures begins to occur for $\Ha_z(y) \approx 0.8 \Ha_{z,c} $ where the flow is dominated by ascending and descending planar jets originating from the sidewalls. The flow structures in this regime are very similar to those observed by \citet{Akhmedagaev:JFM2020} in their simulation data of RBC in a cylindrical cell under strong uniform vertical magnetic field. They can be attributed to quasi-two-dimensional vortex sheets which are often found in magnetohydrodynamic flows with strong magnetic fields~\citep{Zikanov:JFM1998}. As one moves further towards stronger magnetic flux region, the structures originating from the sidewalls extend less into the bulk, until for $\Ha_z(y) \gtrsim \Ha_{z,c}$, the flow is confined only near the sidewalls. These wall-attached flows will be discussed in detail in \S~\ref{sec:WallModes}.


\begin{figure}
  \centerline{\includegraphics[scale = 0.32]{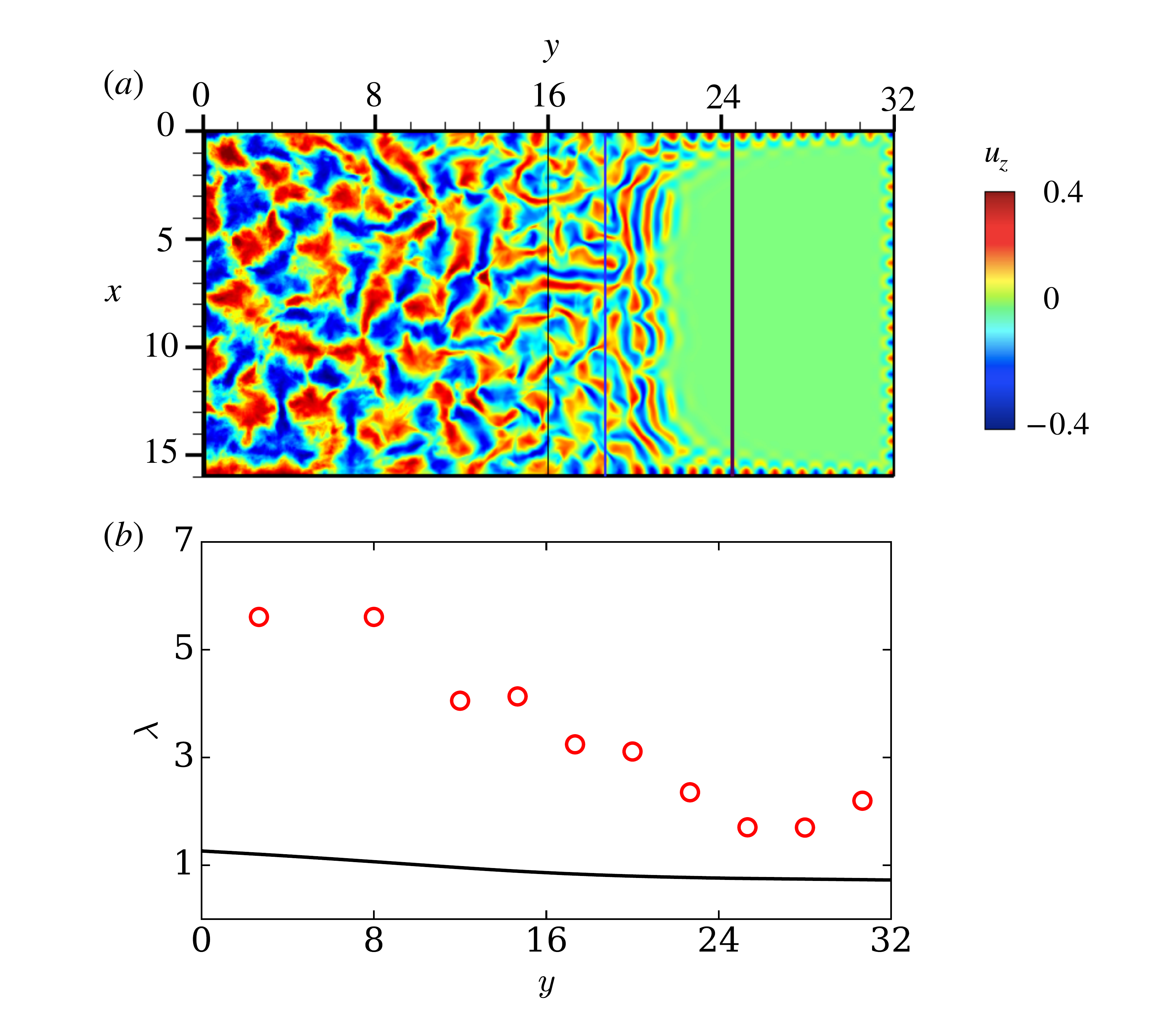}}
  \caption{\textcolor{black}{Pattern analysis for $\delta=9$. (a) Contour plot of time-averaged vertical velocity field on the horizontal midplane. Solid lines are replotted from Fig. \ref{fig:Superstructures}. (b) The evolution of the characteristic horizontal length-scale of the superstructures (red circles) and the critical wavelength $\lambda_c$ (black curve) at the onset of convection along the horizontal $y$-direction.  The length scale of the convection patterns becomes smaller along the direction of the increasing magnetic field and is larger than the critical wavelength.}}
\label{fig:Wavelength}
\end{figure}
It can be visually observed that the \textcolor{black}{size} of the convection patterns decreases along the direction of increasing magnetic field. \textcolor{black}{We focus specifically on the results of $\delta=9$ because the increase of magnetic field along $y$ is gradual in this case, thus enabling us to obtain better statistics for analyzing the variations of the structures with the magnetic field strength. In order to quantitatively analyze the evolution of the size of the structures along $y$, we divide the horizontal midplane into 10 subslices along the $y$-direction and estimate the characteristic length scale of these structures in each subslice using the vertical velocity data. Each subslice spans the entire width of the convection cell. Since the structures are large in the low magnetic flux region, the first two subslices (corresponding to $0<y<5.34$ and $5.34<y<10.67$) are larger than the rest of the subslices; the larger subslices have a width of $5.34$ units and the smaller subslices are $2.67$ units wide.
We employ the procedure outlined in \citet{Pandey:NC2018} to estimate the length scale as elaborated below. For each subslice, we compute the Fourier transform of the vertical velocity field at each snapshot to obtain $\widehat{u}_z(k,\phi_k,t)$, where $k$ is the magnitude of the wavevector and $\phi_k$ is the corresponding azimuthal angle in the wavenumber space. We calculate the azimuthally averaged kinetic energy spectra, which is obtained by
\begin{equation}
    E_u(k,t) = \frac{1}{2\pi} \int_0^{2\pi} |\widehat{u}_z(k,\phi_k,t)|^2d\phi_k,
    \label{eq:Eu_azimuth}
\end{equation}
The above computed kinetic energy spectra are averaged over $20$ snapshots. The wavenumber $k_{max}$ corresponding to which the spectrum is maximum is the characteristic wavenumber\footnote{\textcolor{black}{In case the kinetic energy spectrum has multiple maxima, the characteristic wavenumber is given as the average of the wavenumbers corresponding to the maxima weighted by the kinetic energy contained in these wavenumber shells.}}; and the characteristic length scale $\lambda$ is computed as
\begin{equation}
    \lambda = \frac{2\pi}{k_{max}}.
    \label{eq:Lambda}
\end{equation}
The contour plot of $u_z$ is redrawn in figure~\ref{fig:Wavelength}(a) and the computed characteristic length scale for each subslice is plotted versus $y$ in figure~\ref{fig:Wavelength}(b). Figure~\ref{fig:Wavelength}(b) shows that $\lambda$ decreases with the increasing magnetic field strength along $y$, which is consistent with the visual observation that the size of the patterns decreases along $y$. Note that the characteristic length scale in the weak magnetic flux region is $\lambda=5.5$, which is of the same order as the value of $\lambda \approx 4$ computed by \citet{Pandey:NC2018} using their numerical data for $\Ray=10^5$ and $\Pran=0.021$.
Figure~\ref{fig:Wavelength}(b) also shows that the characteristic length scales computed using our data are larger than the critical wavelength $\lambda_c$~\citep{Chandrasekhar:book:Instability} at the onset of convection for the corresponding Hartmann number.
In the wall-attached convection region where $\Ha_z(y)>\Hazmax$, the computed length scale ranges from $\lambda \approx 1.5$ to $\lambda \approx 2$ and denotes the characteristic size of the wall-modes. These values are fairly close to the analytically derived value of $\lambda_{c,w} = \sqrt{2}$ for wall modes in magnetoconvection with one sidewall~\citep{Busse:PF2008}.}  

\begin{figure}
  \centerline{\includegraphics[scale = 0.35]{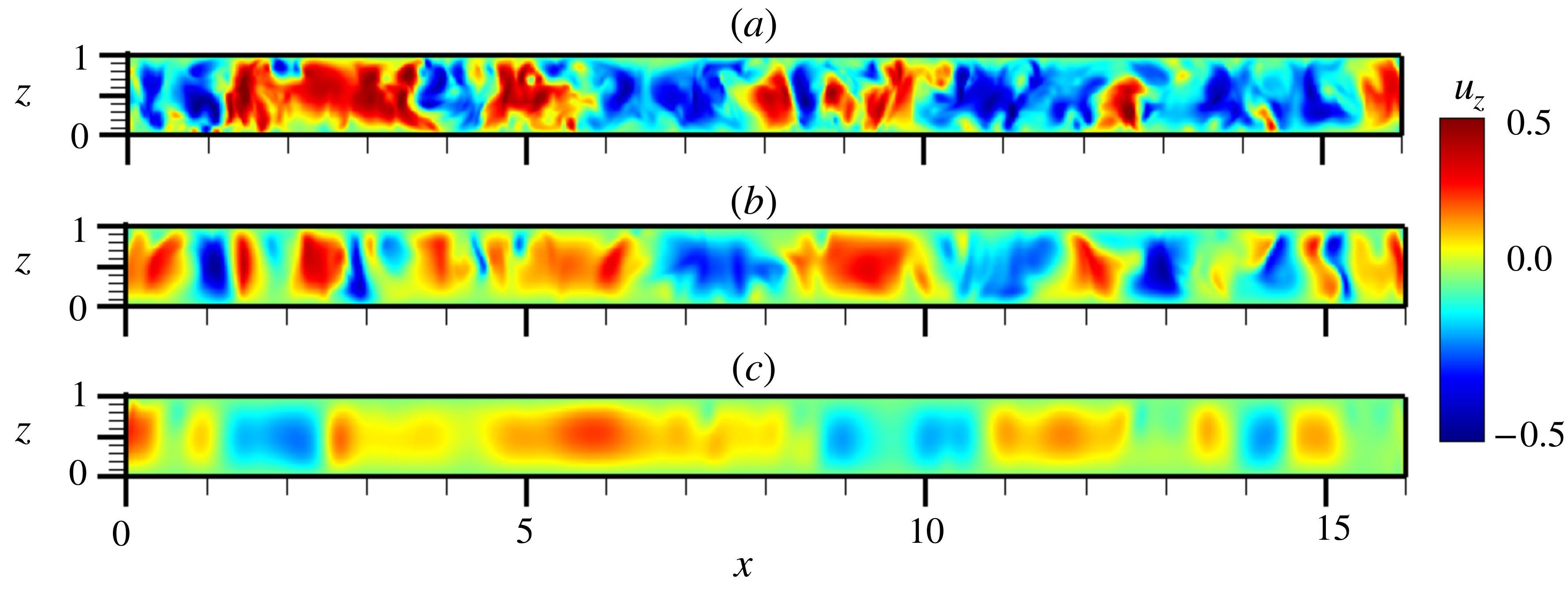}}
  \caption{\textcolor{black}{Contour plots of the time-averaged vertical velocity field for $\delta=9$ in vertical planes at (a) $y=3.4$ ($\Ha_z(y)=25$), (b) $y=11.9$ ($\Ha_z(y)=45$), and (c) $y=19.7$ ($\Ha_z(y)=75$). The velocity fluctuations decrease with the increasing magnetic field strength along $y$.}}
\label{fig:Vertical_structures}
\end{figure}

\begin{figure}
  \centerline{\includegraphics[scale = 0.33]{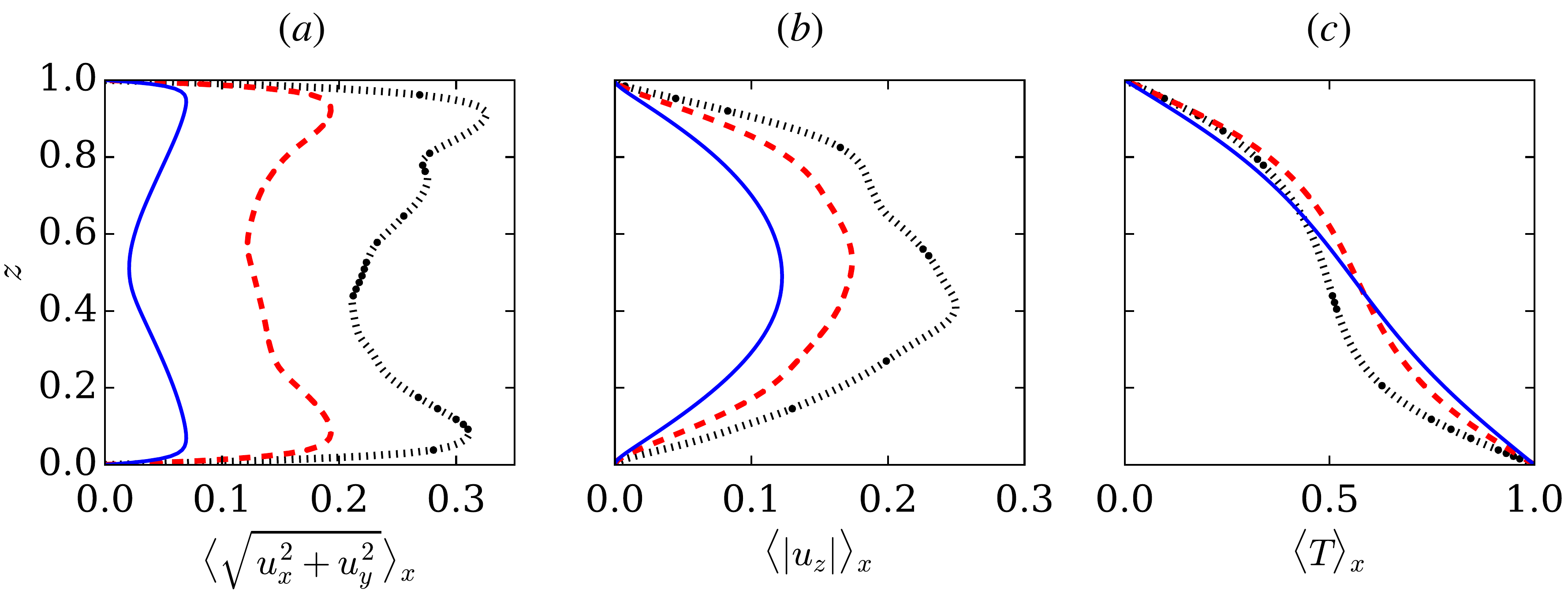}}
  \caption{\textcolor{black}{Profiles for $\delta=9$ of (a) the mean horizontal velocity $\left \langle \sqrt{u_x²+u_y^2} \right \rangle_x$, (b) the mean absolute vertical velocity $\langle |u_z| \rangle_x$, and (c) the mean temperature along the vertical $z$-direction at $y=3.9$ (black dotted curves),  $y=11.9$ (red dashed curves), and $y=19.7$ (blue solid curves). As the magnetic field strength along $y$ increases, the velocity fluctuations decrease and the temperature approaches the linear conduction profile.}}
\label{fig:Vertical_profile}
\end{figure}

\textcolor{black}{We now examine the convection structures on different vertical planes along $y$ for $\delta=9$. In figures~\ref{fig:Vertical_structures}(a,b,c), we show the density plots of vertical velocity on the planes at $y=3.4$, $y=11.9$ and $y=19.7$. The local Hartmann number at these locations are $\Ha_z(y=3.4)=25$, $\Ha_z(y=11.9)=45$, and $\Ha_z(y=19.7)=75$. The velocity fluctuations are strong at $y=3.4$ where the flow is highly turbulent. At $y=11.9$, the flow structures consist of quasi-two-dimensional upward and downward streams occupying the entire plane. The velocity fluctuations become weak at $y=19.7$ and the flow structures become elongated along the $x$-direction, consistent with the observation from figures~\ref{fig:Superstructures}(e) and (j). We plot the mean horizontal velocity $\left \langle \sqrt{u_x^2 + u_y^2} \right \rangle_x$, the mean absolute vertical velocity $\langle |u_z| \rangle_x$, and the mean temperature $\langle T \rangle_x$ versus $z$ on the above three planes in figures~\ref{fig:Vertical_profile}(a,b,c) respectively. These figures further reinstate that the velocity fluctuations decrease along $y$. Moreover, the temperature approaches the linear conduction profile as the local magnetic field strength increases.}

\begin{figure}
  \centerline{\includegraphics[scale = 0.33]{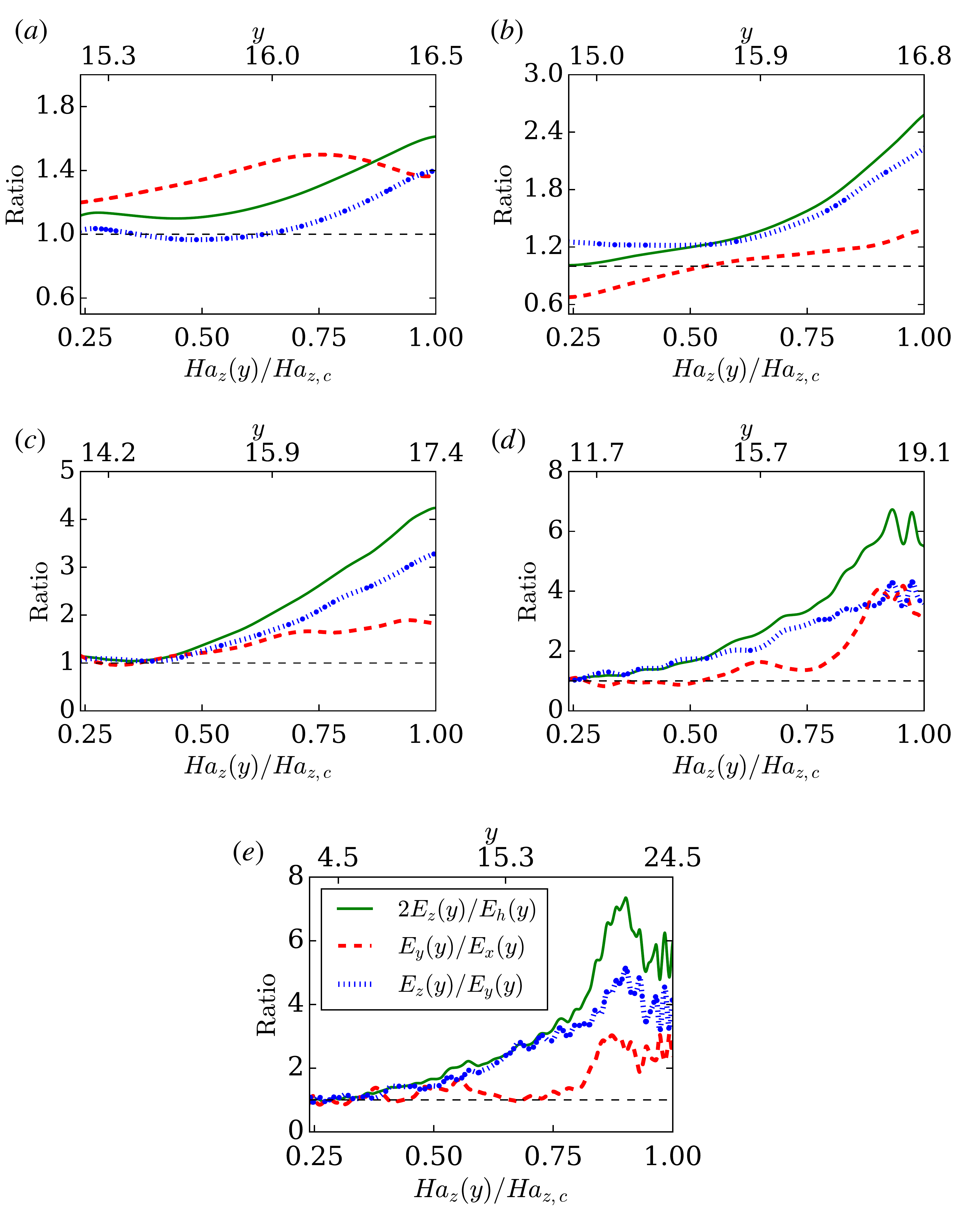}}
  \caption{\textcolor{black}{Mean profiles of the local vertical anisotropy parameter $2E_z(y)/E_h(y)$ (solid green curves), the local horizontal anisotropy parameter $E_y(y)/E_x(y)$ (dashed red curves), and the ratio of the kinetic energy along $z$ and $y$ directions -- $E_z(y)/E_y(y)$ (dotted blue curves) -- versus the normalized Hartmann number, $\Ha_z(y)/\Ha_{z,c}$, based on the mean vertical magnetic field $B_z(y)$, for (a) $\delta=0.01$, (b) $\delta=0.3$, (c) $\delta=1$, (d) $\delta=3$, and (e) $\delta=9$. Both} the vertical and the horizontal anisotropy parameters increase with the local vertical magnetic field strength.}
\label{fig:E_Ha}
\end{figure}

We now quantitatively analyse the isotropy of the flow using our numerical data. Towards this objective, we compute the following \textcolor{black}{ratios}: the local vertical anisotropy parameter $2E_z(y)/E_h(y)$, the local horizontal anisotropy parameter $E_y(y)/E_x(y)$, and the ratio of the kinetic energies along the $z$ and $y$ directions. In the above, $E_z(y) = 0.5 \langle u_z^2 \rangle_{x,z}$, is the vertical kinetic energy, $E_h(y) = 0.5 \langle u_x^2 + u_y^2 \rangle_{x,z}$ is the horizontal kinetic energy, $E_x(y) = 0.5 \langle u_x^2 \rangle_{x,z}$ is the kinetic energy along $x$ direction, and \textcolor{black}{$E_y(y) = 0.5 \langle u_y^2 \rangle_{x,z}$} is the kinetic energy along the $y$ direction, all averaged over the $x$-$z$ plane. To understand the variation of flow anisotropy with magnetic field strength, the aforementioned anisotropy factors are plotted versus \textcolor{black}{$\Ha_z(y)/\Ha_{z,c}$ for different $\delta$ in figures~\ref{fig:E_Ha}(a)--(e).
In these plots, we focus on the regions where $0.25\Ha_{z,c} < \Ha_z(y) < \Ha_{z,c}$ because, 
as discussed later in \S~\ref{sec:GlobalQuantitites}, the flow undergoes significant changes in this parameter interval. 
Further, we add horizontal tick labels with the coordinates of $y$ on the top of each plot to better 
understand the evolution of anisotropy along the horizontal direction.}

\textcolor{black}{The above plots show that} the flow is roughly isotropic in the regions of weak vertical magnetic field corresponding to $\Ha_z(y)<0.3\Ha_{z,c}$. However, \textcolor{black}{with increasing $\Ha_z(y)$,} the vertical velocity fluctuations become more dominant compared to the horizontal ones. This property is consistent with the anisotropy observed by \citet{Yan:JFM2019} and \citet{Akhmedagaev:JFM2020} for strong Hartmann number convection with uniform magnetic fields, and is reminiscent of the stable cellular regime described by \citet{Zuerner:JFM2020}.  
\textcolor{black}{The increase of anisotropy with $\Ha_z(y)$ becomes more pronounced as the fringe-width, governed by $\delta$, increases. This is because as the fringe-width increases, the span of $y$ falling in the regime of $0.25\Ha_{z,c} < \Ha_z(y) < \Ha_{z,c}$ also increases as exhibited in figures~\ref{fig:E_Ha}(a)--(e). Thus, we get better statistics, and hence, clearer trends in this regime for large $\delta$.} 

\textcolor{black}{Figures~\ref{fig:E_Ha}(a)--(e) also show that }for strong magnetic fields, there is anisotropy among the horizontal components of velocity as well. For $\Ha_z(y)> 0.65 \Ha_{z,c}$, $E_y(y)$ begins to dominate $E_x(y)$, implying that the velocity fluctuations in the $y$ direction are stronger compared to that in the $x$ direction. This is also consistent with figures~\ref{fig:Superstructures}(a)--(j) which show that for moderately strong Hartmann numbers, the superstructure rolls orient themselves perpendicular to the longitudinal sidewalls. 
In the next subsection, we discuss the effects of the fringing magnetic fields on the local as well as global momentum and heat transport.

\subsection{Heat and momentum transport}\label{sec:GlobalQuantitites}
We analyze the spatial variation of the large-scale heat and momentum transport in our numerical setup. Towards this objective, we first compute the local planar Reynolds number $\Rey(y)$ and Nusselt number $\Nu(y)$ for every $\delta$. These are given by
\begin{equation}
    \Rey(y) = \sqrt{\frac{\Ray}{\Pran}}\langle u_x^2 + u_y^2 + u_z^2 \rangle_{x,z,t_N}^{1/2},
    \quad
    \Nu(y) = 1 + \sqrt{\Ray \Pran}\langle u_z T \rangle_{x,z,t_N}. 
\end{equation}
It must be noted that in order to avoid clutter, \textcolor{black}{$\Nu(y)$} and $\Rey(y)$ are computed for every 120th $x$-$z$-plane.  Moreover, $\Nu(y)$ exhibits strong spatial fluctuations even after averaging over all the timeframes; these fluctuations are smoothened by computing the moving average of $\Nu(y)$ using a window-size of 5 $x$-$z$-planes. 

\begin{figure}
  \centerline{\includegraphics[scale = 0.35]{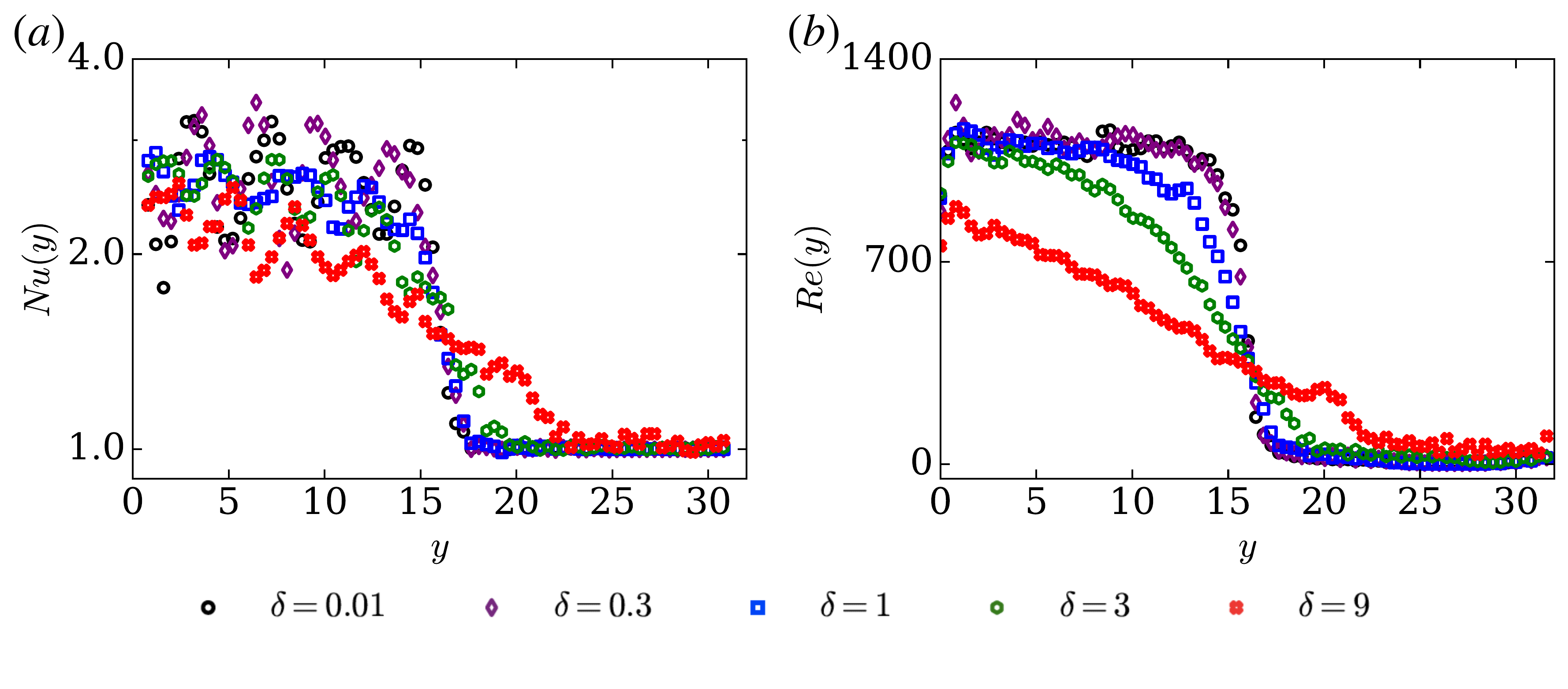}}
  \caption{\textcolor{black}{Variations of the \textcolor{black}{local planar} Nusselt and Reynolds numbers with the longitudinal direction $y$ . For $\delta=0.01$ to $9$: (a) local Nusselt number, ${\Nu}(y)$, and (b) local Reynolds number, ${\Rey}(y)$, versus $y$.}} 
\label{fig:y_ReNu_continuous}
\end{figure}

\textcolor{black}{We plot $\Nu(y)$ and $\Rey(y)$ versus $y$ in figures~\ref{fig:y_ReNu_continuous}(a) and (b) respectively. Figure~\ref{fig:y_ReNu_continuous}(a) shows that $\Nu(y)$ fluctuates between 2.0 and 4.0 in the region of weak magnetic flux, drops steeply in the fringe-zone and becomes close to unity in the strong magnetic flux region. Figure~\ref{fig:y_ReNu_continuous}(b) shows that $\Rey(y)$ follows a similar trend; it is large in the weak magnetic flux region, drops steeply in the fringe-zone, and becomes negligibly small in the strong magnetic flux region. The gradients of both $\Rey(y)$ and $\Nu(y)$ curves in the fringe-zone decrease with an increase of $\delta$. This is expected because $B_z$, which suppresses kinetic energy and heat transport in the fluid, has steeper gradients for small $\delta$.} 

\begin{figure}
  \centerline{\includegraphics[scale = 0.35]{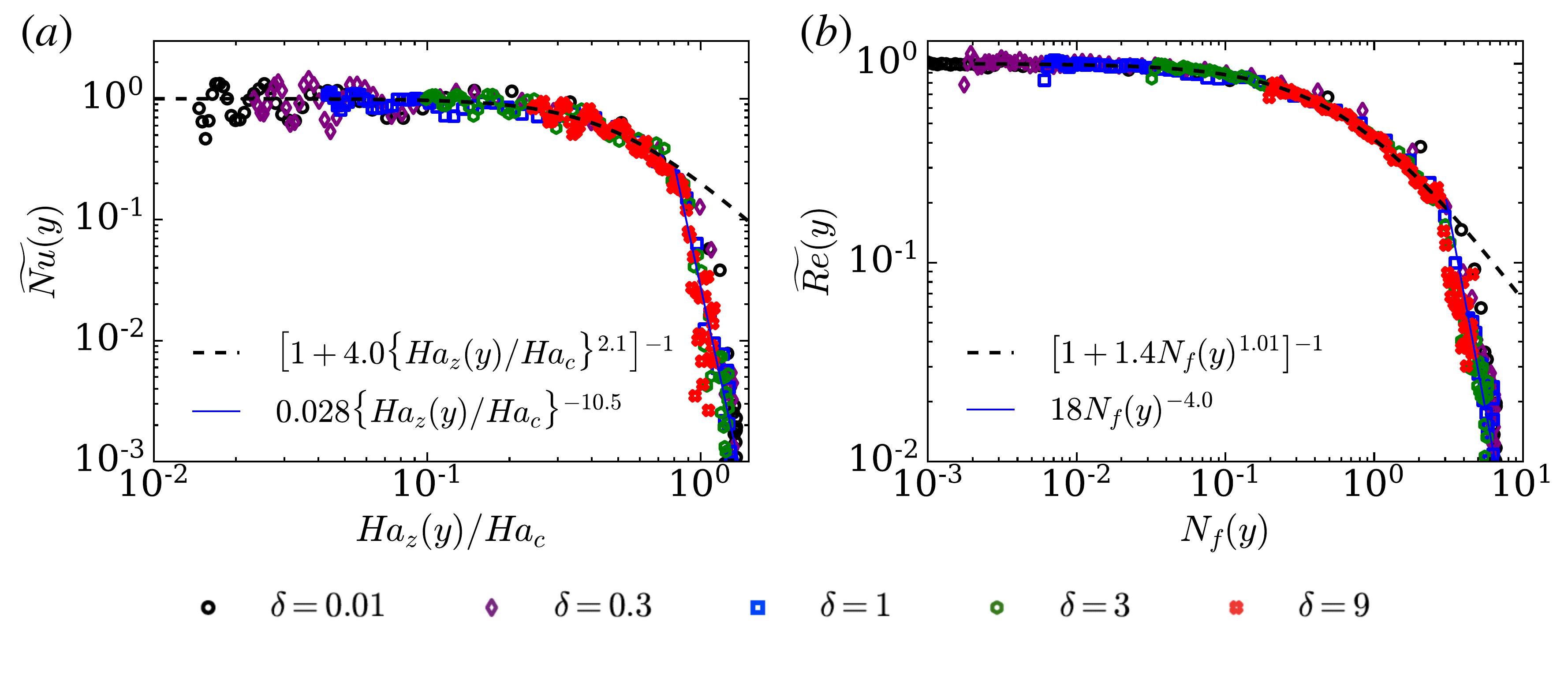}}
  \caption{Variations of the \textcolor{black}{normalized local} Nusselt and Reynolds numbers with the \textcolor{black}{normalized} local vertical magnetic field strengths. For $\delta=0.01$ to $9$: (a) normalized local Nusselt number, $\widetilde{\Nu}(y)$, versus the normalized local Hartmann number, $\Ha(y)/Ha_c$, based on vertical magnetic field strength; and (b) normalized local Reynolds number, $\widetilde{\Rey}(y)$, versus the local free-fall interaction parameter, $N_{f}(y)$. The best-fit curves for the above data are also shown.} 
\label{fig:Ha_ReNu_continuous}
\end{figure}

\textcolor{black}{We now look for a universal dependence of the local heat and momentum transport on the local magnetic field strength.} Taking inspiration from the recent experimental work of \citet{Zuerner:JFM2020} on thermal convection under uniform vertical magnetic fields, we plot the normalized local Nusselt number $\widetilde{\Nu}(y)$ versus the normalized local vertical Hartmann number $\Ha_z(y)/\Ha_{z,c}$. The normalized local Nusselt number is given by
\begin{equation}
    \widetilde{\Nu}(y) = \frac{\Nu(y)-1}{\Numax-1},
    \label{eq:Nu_normalized}
\end{equation}
where $\Numax=2.66$ (see \S~\ref{sec:Numerical_method}). Figure~\ref{fig:Ha_ReNu_continuous}(a) shows that for all $\delta$, the points collapse  well into a single curve, thus showing a universal dependence of $\Nu(y)$ on $\Ha_z(y)$. The plot shows that for $\Ha_z(y)\lesssim 0.3\Ha_{z,c}$, the Nusselt number does not change significantly and remains close to its maximum value. This regime corresponds to the turbulent and isotropic regime. For $\Ha_z(y)\gtrsim 0.3\Ha_{z,c}$, the flow transitions into the cellular regime \citep[as per the nomenclature of][]{Zuerner:JFM2020} and the local Nusselt number starts to drop sharply as $\Ha_z(y)$ increases. For $\Ha_z(y)<0.85 \Ha_{z,c}$, the best fit curve for our data is
\begin{equation}
\widetilde{\Nu}(y) = \left[1 + \chi_1 \left\{\frac{\Ha_z(y)}{\Ha_{z,c}} \right\}^{\gamma_1} \right]^{-1},
\label{eq:Nu_fit_1}
\end{equation}
\textcolor{black}{where $\gamma_1 = 2.1 \pm 0.3$ and $\chi_1 = 4.0 \pm 0.7$. Interestingly, the values of $\chi_1$ and especially $\gamma_1$ are close to those observed by \citet{Zuerner:JFM2020}, who reported $\gamma_1 = 2.03 \pm 0.06$ and $\chi_1 = 5.9 \pm 0.3$ for uniform magnetic fields. }
For $\Ha_z(y) > 0.85 \Ha_{z,c}$, $\tilde{\Nu}(y)$ decreases more steeply than (\ref{eq:Nu_fit_1}) and is described by the following power law,
\begin{equation}
    \widetilde{\Nu}(y) = 0.028 \left\{\frac{\Ha_z(y)}{\Ha_{z,c}} \right\}^{-10.5\pm 0.9}\,.
    \label{eq:Nu_fit_2}
\end{equation}
This regime corresponds to the wall-mode regime in which the flow gets further suppressed and begins to confine itself near the sidewalls. It must be noted that the \citet{Zuerner:JFM2020} could not obtain a best-fit expression for $\Nu$ in the wall-mode regime due to lack of data close to the wall.

We conduct a similar analysis for $\Rey(y)$ as well. We compute the normalized Reynolds number given by 
\begin{equation}
    \widetilde{\Rey}(y) = \frac{\Rey(y)}{\Remax}, 
    \label{eq:Re_normalized}
\end{equation}
where $\Remax=1115$ (see \S~\ref{sec:Numerical_method}). The results of \citet{Zuerner:JFM2020} suggest that $\widetilde{\Rey}(y)$ should scale as $\Ha_z(y)/\sqrt{\Ha_{z,c}}$ instead of $\Ha_z(y)/\Ha_{z,c}$. \textcolor{black}{Let us now consider the local free-fall interaction parameter $N_{f}(y)$, which is given by
\begin{equation}
    N_{f}(y) = \Ha^2_z(y) \sqrt{\frac{\Pran}{\Ray}}.
    \label{eq:Nf}
\end{equation}
}\textcolor{black}{For large Rayleigh numbers,  the critical Hartmann number can be approximated as $\Ha_{z,c} \approx \sqrt{\Ray}/\pi$~\citep{Chandrasekhar:book:Instability}. Using this relation, we get} 
\begin{equation}
    N_{f}(y) = \Ha^2_z(y) \frac{\pi}{\sqrt{\Ray}} \frac{\sqrt{\Pran}}{\pi} \approx 
    \frac{\Ha^2_z(y)}{\Ha_{z,c}} \frac{\sqrt{\Pran}}{\pi} \sim  \frac{\Ha^2_z(y)}{\Ha_{z,c}}.
    \label{eq:Nf}
\end{equation}
The above relation shows that $N_f(y)$ is proportional to ${\Ha^2_z(y)}/{\Ha_{z,c}}$, implying that $\widetilde{\Rey}(y)$ scales with $N_f(y)$. Thus, we plot $\widetilde{\Rey}(y)$ versus $N_f(y)$. The plot in figure~\ref{fig:Ha_ReNu_continuous}(b) shows that, similar to the Nusselt number, the data points for all $\delta$ collapse into a single curve. For $N_{f}(y) < 0.17$, which corresponds to $\Ha_z(y) < 0.22 \Ha_{z,c}$, the Reynolds number retains its maximum value ($\Rey_{max}$). At higher $N_{f}(y)$, the Reynolds number starts to fall sharply with the interaction parameter. For $N_{f}(y) < 3.2$, which corresponds to $\Ha_z(y) < 0.95 Ha_{z,c}$, the best fit curve for our data is given by
\begin{equation}
\widetilde{\Rey}(y) = \left[1 + \chi_2 \left\{N_{f}(y) \right\}^{\gamma_2} \right]^{-1},
\label{eq:Re_fit_1}
\end{equation}
with $\chi_2 = 1.40 \pm 0.03$ and $\gamma_2 = 1.01 \pm 0.03$. \textcolor{black}{The value of $\gamma_2$ is again close to that observed by \citet{Zuerner:JFM2020}, who reported $\gamma_2 = 0.87 \pm 0.03$ for uniform magnetic fields.} For  $\Ha_z(y)>0.95 \Ha_{z,c}$, which corresponds to the wall-attached convection regime, $\widetilde{\Rey}(y)$ falls even more steeply than (\ref{eq:Re_fit_1}) and is described by the following power law,
\begin{equation}
\widetilde{\Rey}(y) = 18\{N_{f}(y)\}^{-4.0\pm 0.2}\,.
\label{eq:Re_fit_2}
\end{equation}

\begin{figure}
  \centerline{\includegraphics[scale = 0.33]{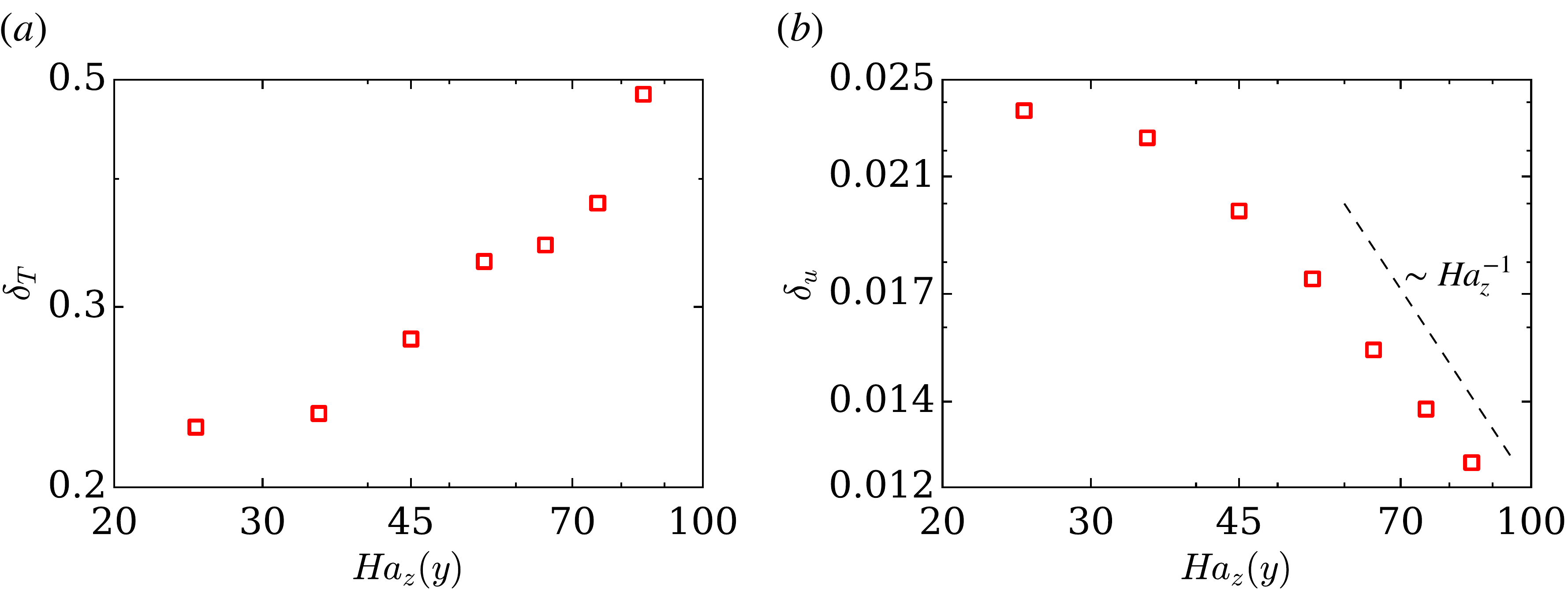}}
  \caption{\textcolor{black}{Boundary analysis for $\delta=9$. (a) Local thermal boundary layer thickness $\delta_T(y)$ and (b) local viscous boundary layer thickness $\delta_u(y)$ near the top and bottom walls  versus the local Hartmann number $\Ha_z(y)$. The thermal boundary layer thickness increases with the increase of $\Ha_z(y)$. The viscous boundary layer thickness decreases with increasing $\Ha_z(y)$ and approaches Hartmann type scaling of $\delta_u \sim \Ha_z^{-1}$ at large Hartmann numbers. }}
\label{fig:BL_Ha}
\end{figure}

\textcolor{black}{We now analyze the impact of the local heat and momentum tranport on the evolution of the viscous and thermal boundary layers near the top and bottom walls of our convection cell. Again, we concentrate on $\delta=9$ to obtain better statistics and a clearer picture of the evolution. The thermal boundary layer thickness $\delta_T$ is the depth where a linear fit of the mean temperature profile near the wall intersects with the mean bulk temperature ($T=0.5$). The viscous boundary layer thickness is the depth where a linear fit of the horizontal velocity ($u_h = \langle (u_x^2+u_y^2)^{1/2} ~\rangle_x$) profile near the wall intersects with its local maximum. See \citet{Breuer:PRE2004}, \citet{Ahlers:RMP2009} and \citet{Scheel:JFM2012} for a detailed procedure to compute the boundary layer thicknesses. We compute the thermal and viscous boundary layer thicknesses at different $y$-coordinates for $\delta=9$ and plot them versus the corresponding local Hartmann number $\Ha_z(y)$ in figures~\ref{fig:BL_Ha}(a,b), respectively.}

\textcolor{black}{Figure~\ref{fig:BL_Ha}(a) shows that the thermal boundary layer thickness increases with $\Ha_z(y)$. This is expected because the thermal boundary layer thickness is inversely proportional to the local Nusselt number $\Nu(y)$ which decreases with increasing $\Ha_z(y)$. The viscous boundary layer, on the other hand, exhibits an interesting behaviour.  As evident in figure~\ref{fig:BL_Ha}(b), the boundary layer thickness $\delta_u$ decreases with $\Ha_z(y)$, with the decrease becoming steeper at large $\Ha_z(y)$. To gain insight into this behavior, we take inspiration from the work of \citet{Lim:JFM2019} and using dimensional analysis on the momentum equation, we obtain the following scaling for $\delta_u$:
\begin{equation}
    \delta_u \sim \frac{1}{\sqrt{\Rey(y) + \Ha_z^2(y)}}
    \label{eq:BL_scaling}
\end{equation}
The above scaling suggests that for $\Ha_z(y) \ll \sqrt{\Rey(y)}$, $\delta_u$  follows the Prandtl-Blasius type scaling of $\delta_u \sim \Rey^{-1/2}$ in which the boundary layer thickness should increase with the decrease of $\Rey(y)$, and hence, with the increase of $\Ha_z(y)$. However, we do not observe such regime in our convection cell. When $\Ha_z(y) \sim \sqrt{\Rey(y)}$, the boundary layer thickness varies marginally with $y$, as observed for $\Ha_z(y) < 50$ from our data. For $\Ha_z(y) \gg \sqrt{\Rey(y)}$, the boundary follows the Hartmann type scaling of $\delta_u \sim \Ha_z^{-1}$ and hence, $\delta_u$ decreases sharply with $\Ha_z(y)$. From figure~\ref{fig:BL_Ha}(b), it can be seen that $\delta_u$ approaches the Hartmann type scaling at $\Ha_z(y) \approx 70$. 
}

\begin{figure}
  \centerline{\includegraphics[scale = 0.35]{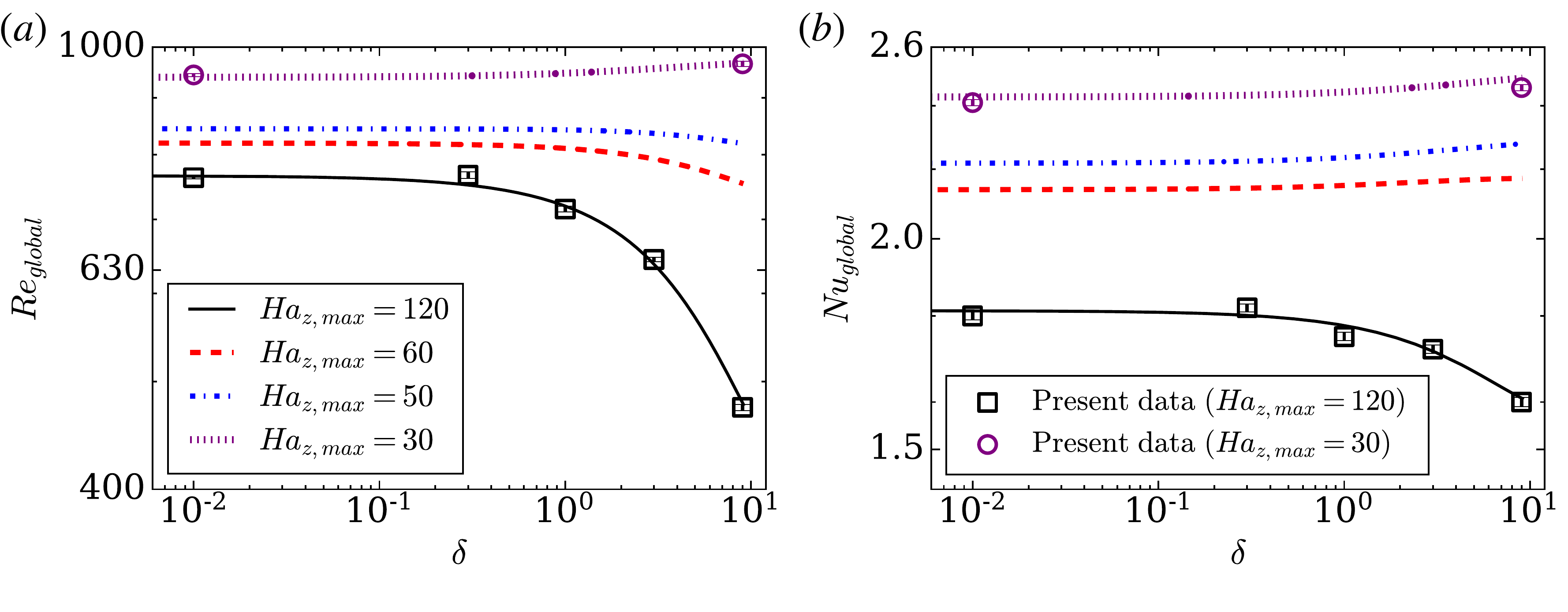}}
  \caption{Plots of (a) the global Reynolds number $\Reglobal$ (black squares for \textcolor{black}{$\Hazmax=120$ and purple circles for $\Hazmax=30$}), computed from our numerical data using equation~(\ref{eq:ReDirect}), versus $\delta$, and (b) the global Nusselt number $\Nuglobal$ (black squares \textcolor{black}{for $\Hazmax=120$ and purple circles for $\Hazmax=30$}), computed from our numerical data using equation~(\ref{eq:NuDirect}), versus $\delta$. Also shown in the plots are the estimates (solid curves) of $\Reglobal$ and $\Nuglobal$ computed using equations~\eqref{eq:Reglobal} and \eqref{eq:Nuglobal} for different values of $\Hazmax$.}
\label{fig:NuRe_diffHa}
\end{figure}

Having studied the local variations of heat and momentum transport \textcolor{black}{along with their associated boundary layers}, we now analyze the global  Reynolds  and Nusselt numbers ($\Reglobal$ and $\Nuglobal$ respectively) and their dependence on $\delta$. These global quantities are computed numerically using 
equations~(\ref{eq:ReDirect}) and (\ref{eq:NuDirect}) from our simulation data and are plotted versus $\delta$ in figures~\ref{fig:NuRe_diffHa}(a) and (b) as black squares. The figures show that both $\Reglobal$ and $\Nuglobal$ decrease as $\delta$ increases. This result is counter intuitive because for large $\delta$, bulk convection is completely ceased only in a small region (as seen in figure~\ref{fig:Superstructures}) and hence one would expect the overall heat and momentum transport to be stronger. However, a careful look at figure~\ref{fig:Bz_By_y}(a) reveals that the vertical magnetic field is stronger in the weak magnetic flux region for large fringe-widths, which, in turn, weakens convection in that region. The increase in convection in the strong magnetic flux region is unable to compensate the suppression of convection in the weak magnetic flux region, resulting in a decrease of overall heat and mass transport for large fringe-widths. 

The above results, however, do not indicate whether such a variation of global Reynolds and Nusselt numbers with $\delta$ holds for all $\Hazmax$. To explore this aspect, we need to estimate $\Reglobal$ and $\Nuglobal$ for a given value of $\Hazmax$ and $\delta$. Towards this objective, we make use of the best fit relations for $\Nu(y)$ and $\Rey(y)$ given by equations~(\ref{eq:Nu_fit_1}), (\ref{eq:Nu_fit_2}), (\ref{eq:Re_fit_1}), and (\ref{eq:Re_fit_2}) obtained earlier, and numerically integrate them over the entire domain to estimate global Reynolds and Nusselt numbers as follows:
\begin{eqnarray}
\Reglobal &=& \frac{1}{l_y}\int_0^{l_y} \Rey(y)dy = \frac{1}{l_y}\int_0^{l_y} \Rey(\Ha_z(y))dy, \label{eq:Reglobal}\\
\Nuglobal &=& \frac{1}{l_y}\int_0^{l_y} \Nu(y)dy = \frac{1}{l_y}\int_0^{l_y} \Nu(\Ha_z(y))dy, \label{eq:Nuglobal}
\end{eqnarray}
where we recall that
\begin{equation*}
    \Ha_z(y) = \Hazmax \frac{\langle B_z \rangle_{x,z}}{\Bzmax},
\end{equation*}
and $B_z=B_z(x,y,z,\delta)$ is given by equation~(\ref{eq:Bz}). Thus, for given $\Hazmax$ and $\delta$, one can estimate $\Reglobal$ and $\Nuglobal$ using equations~(\ref{eq:Reglobal}) and (\ref{eq:Nuglobal}).  We compute these estimates for $\Hazmax=120$, 60, 50, and 30, and for $\delta$ ranging from 0.005 to 9.1 in increments of 0.01. 
The estimated values of $\Reglobal$ and $\Nuglobal$ are plotted versus $\delta$ in figures~\ref{fig:NuRe_diffHa}(a) and (b) as solid curves. The curves for $\Hazmax=120$ fit the data points of directly computed $\Reglobal$ and $\Nuglobal$ (black squares) very well, implying that equations~(\ref{eq:Reglobal}) and (\ref{eq:Nuglobal}) provide fairly accurate estimates. \textcolor{black}{In the process of computing the estimates of $\Reglobal$ and $\Nuglobal$ for $\Hazmax < 120$, we assume that the equations and coefficients given in relations~(\ref{eq:Nu_fit_1}), (\ref{eq:Nu_fit_2}), (\ref{eq:Re_fit_1}), and (\ref{eq:Re_fit_2}) remain unchanged as the maximum Hartmann number is varied. We also assume that the values of $\Ha_z(y) = 0.85 \Ha_{z,c}$ and $N_f(y)=3.2$, which were observed to separate the two components of the piecewise functions for $\Nu(y)$ and $\Rey(y)$ for $\Hazmax=120$, remain the same for all values of $\Hazmax$.  These assumptions are based on the fact that the dependence of $\Nu(y)$ and $\Rey(y)$ on $\Ha_z(y)$ is suggested to be universal since the normalized local Nusselt and Reynolds number curves for different $\delta$ collapse as shown in figures~\ref{fig:Ha_ReNu_continuous}(a,b).} 

Figure~\ref{fig:NuRe_diffHa}(a) also shows that the decrease of $\Reglobal$ with $\delta$ becomes less apparent with a decrease of $\Hazmax$. In fact, for $\Hazmax=30$, $\Reglobal$ increases marginally with $\delta$. Figure~\ref{fig:NuRe_diffHa}(b) indicates that $\Nuglobal$ increases with $\delta$ for $\Hazmax=60$, 50, and 30, with the above increase becoming more apparent with the decrease of $\Hazmax$. \textcolor{black} {To test the correctness of the above predictions, we conduct two more direct numerical simulations of our magnetoconvection setup for $\Hazmax=30$ and $\delta=0.01$ and 9. The grid-size and the numerical schemes are the same as those described in \S~\ref{sec:Numerical_method}. We run the simulations for 10 free-fall time units after attaining a statistically steady state. We compute the time-averaged global Reynolds and Nusselt numbers using the above two simulation data and plot them in figures~\ref{fig:NuRe_diffHa}(a) and (b) respectively. The figures show that the simulation results are in close agreement with the predictions, thereby reinforcing our theory.} It is therefore clear that for smaller maximum magnetic field strengths, the overall heat and momentum transport increases with the increase of fringe-width. This implies that for small values of $\Hazmax$, the suppression of convection in the weak magnetic flux region is small compared to the increase of convection in the strong magnetic flux region. Thus, we can conclude that the variation of $\Reglobal$ and $\Nuglobal$ with the fringe-width depends critically on $\Hazmax$. Finally, we discuss the dynamics of wall-attached convection in the strong magnetic flux region in the next subsection.

\subsection{Wall-attached convection}\label{sec:WallModes}
\begin{figure}
  \centerline{\includegraphics[scale = 0.38]{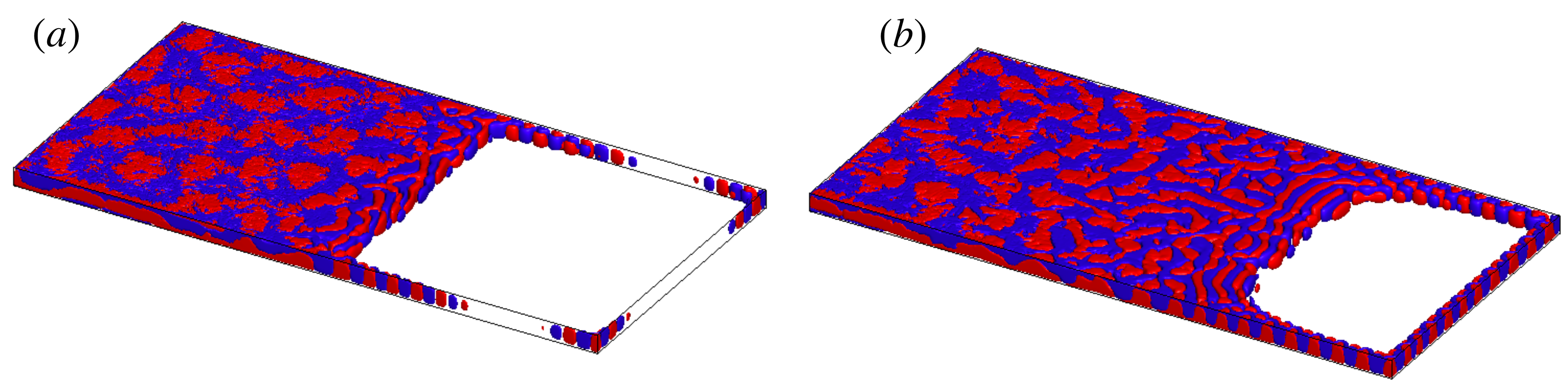}}
  \caption{Isosurfaces of $u_z=0.01$ (red) and $u_z=-0.01$ (blue) for (a) $\delta=1$ and (b) $\delta=9$. In the regions with strong magnetic fields, convective motions are confined near the sidewalls.}
\label{fig:Wall_modes}
\end{figure}

We plot the time-averaged isosurfaces of $u_z = 0.01$ (red) and $u_z = -0.01$ (blue) for $\delta=1$ and $\delta=9$ in figures~\ref{fig:Wall_modes}(a) and (b), respectively. The figures show that in the strong magnetic flux region, the flow in the bulk is completely suppressed with alternating up-and downwelling flow regions attached to the sidewalls. The structure of the wall-modes is consistent with that observed by \citet{Liu:JFM2018} in their simulations of thermal convection in a rectangular domain with uniform vertical magnetic field. However, unlike for rotating convection \citep{Horn:JFM2017,Zhang:PRL2020} or convection in cylindrical domains \citep{Akhmedagaev:JFM2020}, the wall modes do not oscillate or move along the sidewalls \citep{Grannan:JFM2022,Schumacher:JFM2022}. 

The wall-modes are dense and uniformly distributed along the sidewalls for $\delta=9$. However, for $\delta=1$, they are visibly thinner with isosurfaces clustering more towards the corners instead of being uniformly distributed along the sidewalls. The clustering occurs because of weaker Lorentz force in the corners, which, in turn, results due to the electric current getting further constrained by the two insulated sidewalls in proximity of each other. The above trend in the local differences of the wall modes is also observed for $\delta=0.01$, 0.3, and 3 as shown later in this section.

\begin{figure}
  \centerline{\includegraphics[scale = 0.38]{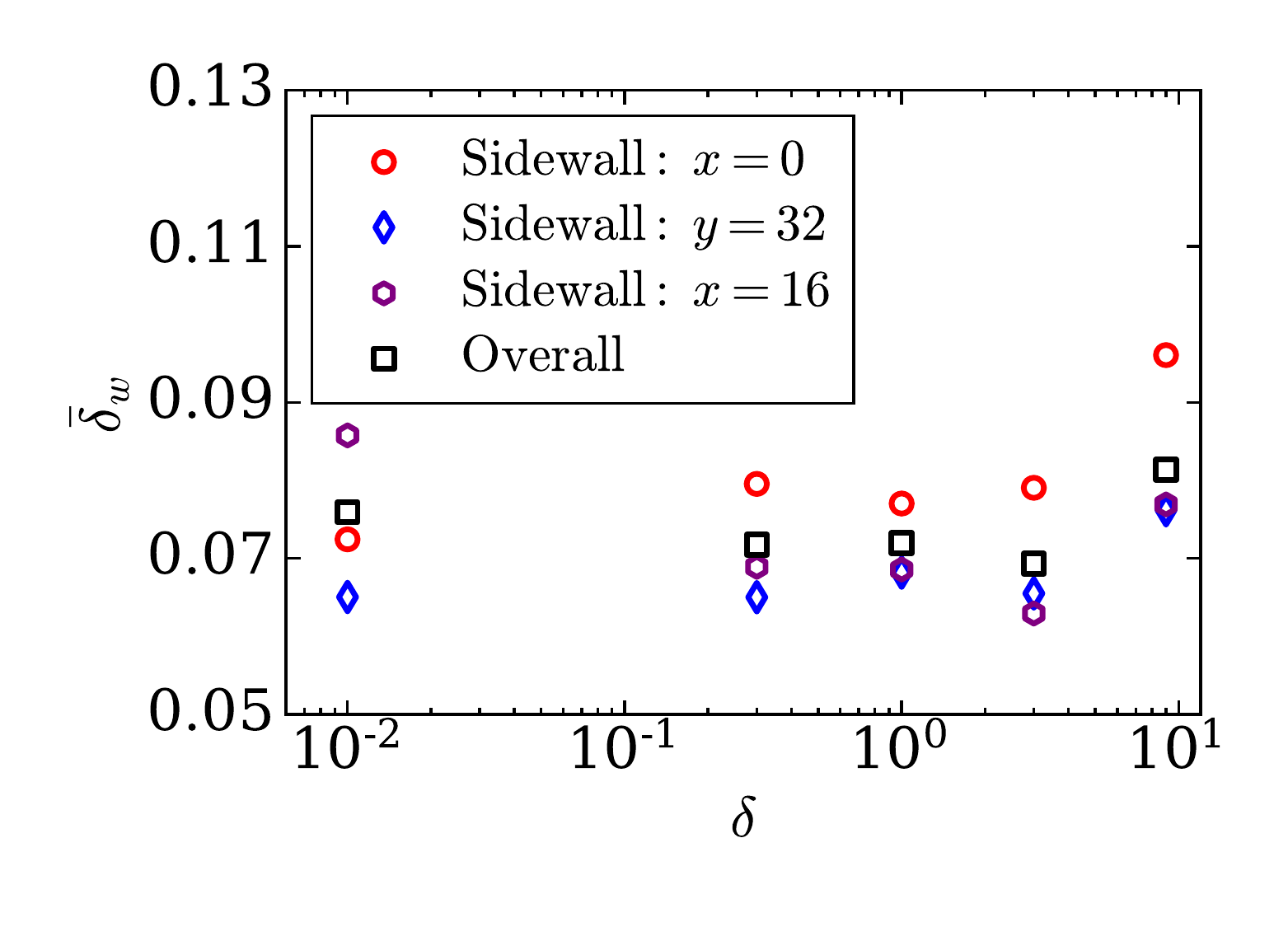}}
  \caption{The average distance between the wall modes and the sidewalls versus $\delta$ in the regions of $\Ha_z(y) > \Ha_{z,c}$. The wall modes are more attached to the lateral sidewalls than the longitudinal sidewalls. }
\label{fig:Wall_mode_distance}
\end{figure}

For each sidewall in the strong magnetic flux region, we compute the mean distance $\bar{\delta}_w$ of the wall modes from the sidewall. This distance is computed as follows,
\begin{equation}
    \bar{\delta}_w = \frac{\int {\delta_w u_{z,max}}dl}{\int {u_{z,max}}dl},
\end{equation}
where $dl = dy$ for the longitudinal sidewall and $dl = dx$ for the lateral sidewall. In the above, $\delta_w$ is the shortest distance between the sidewall and the point of maximum absolute velocity $u_{z,max}$ adjacent to the sidewall, and the summations are over the regions where $\Ha_z(y)>\Ha_{z,c}$. Thus, $\bar{\delta}_w$ is the weighted average of the shortest distance between the point of maximum absolute velocity and the sidewall, with the maximum absolute velocity being the weights. We also compute the mean distance averaged over all the sidewalls. We plot the mean distances for each sidewall as well as the overall mean distance versus $\delta$ in figure~\ref{fig:Wall_mode_distance}. The figure shows that the wall modes are in general more attached to the lateral sidewall ($y=32$)  compared to the longitudinal sidewalls ($x=0$ and $x=16$). There is, however, no clear trend regarding the variation of $\bar{\delta}_w$ with the fringe-width.

\begin{figure}
  \centerline{\includegraphics[scale = 0.38]{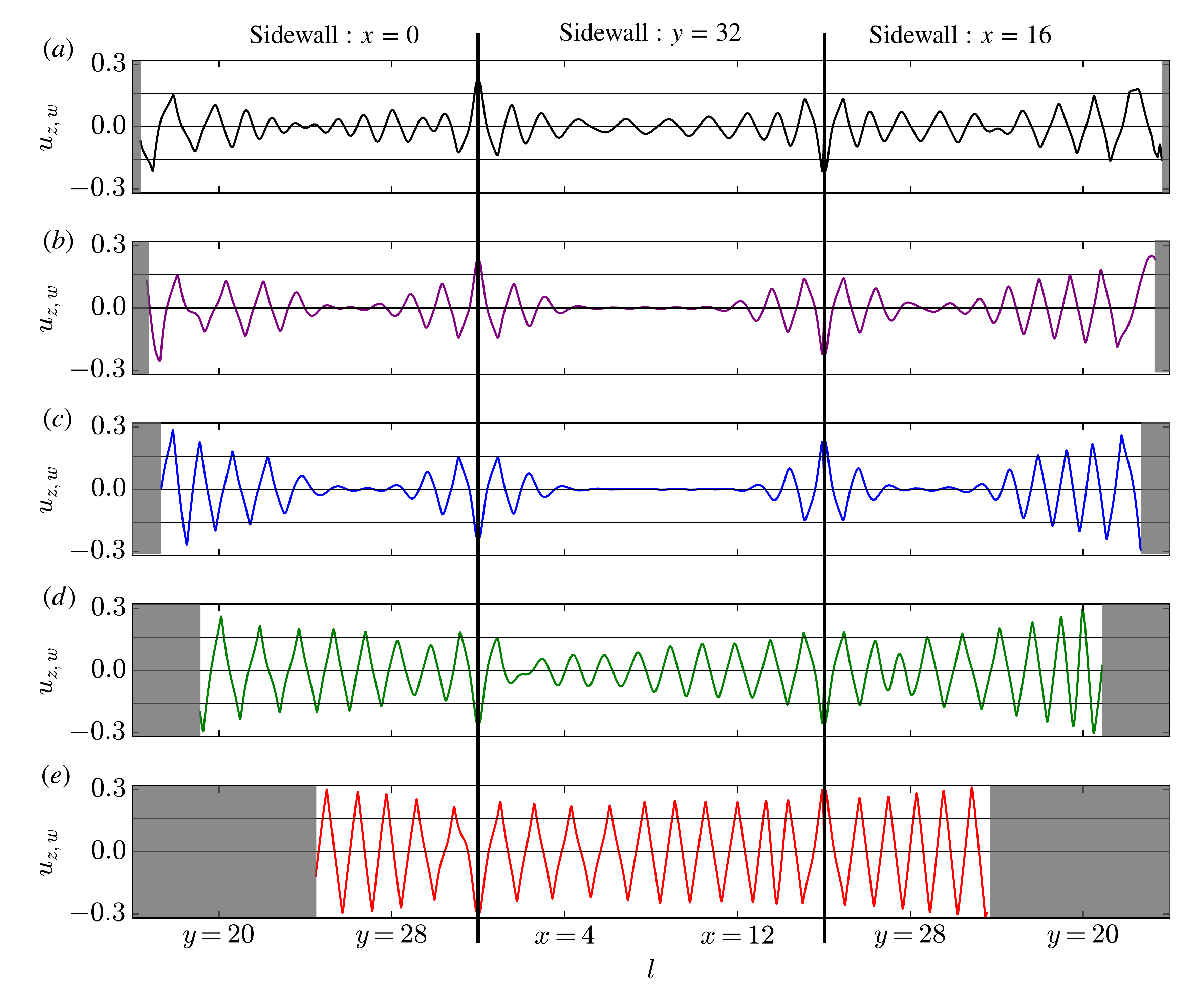}}
  \caption{ Vertical velocity $u_{z,w}$ measured on the horizontal midplane at distance of $\bar{\delta}_w$ from the sidewalls in the regions of $\Ha(y)>\Ha_{z,c}$ (nonshaded regions) for (a) $\delta=0.01$ (black curve), (b) $\delta=0.3$ (purple curve), (c) $\delta=1$ (blue curve), (d) $\delta=3$ (green curve), and (e) $\delta=9$ (red curve). The shaded regions correspond to $\Ha_z(y) < \Ha_{z,c}$ where the flow is not fully suppressed in the bulk. The amplitudes of the wall modes in the regions away from the corners exhibit a non-monotonic dependence on $\delta$. Note that the above plots are resolved by a total of at least 9000 grid points in the horizontal direction.}
\label{fig:Wall_mode_amplitude}
\end{figure}

We further analyze the amplitudes of the wall modes along the three sidewalls for our runs. Towards this objective, we measure the vertical velocity field at every point on the horizontal midplane at a distance of $\bar{\delta}_w$ from the sidewalls. We plot this vertical velocity profile along every sidewall in figure~\ref{fig:Wall_mode_amplitude}. The figure shows that with the exception of $\delta=9$, the amplitudes of the wall modes are stronger near the corners and become weaker as one moves away from the corners. This behavior in the amplitudes is consistent with figure~\ref{fig:Wall_modes}(a) which shows that the wall-modes are clustered in the corners. As explained earlier, these local differences in amplitudes occur due the current being more constrained in the corners resulting in weaker convection-suppressing Lorentz force in these regions. Figure~\ref{fig:Wall_mode_amplitude} also shows that near the corners, the amplitudes of the wall modes grow with an increase of $\delta$. However, in the regions away from the corners, the amplitudes decrease as $\delta$ is decreased from 9 to 1; and then increases with a further decrease of $\delta$. For $\delta=0.3$ and 1, the amplitudes become negligible small away from the corners. However, for $\delta=0.01$, although the amplitudes of the modes away from the corners are smaller than those near the corners, they are larger than the amplitudes of wall modes away from the corners for $\delta=0.3$ and 1.

The reason for the non-monotonic behavior of wall-modes with $\delta$ away from the corners is not yet clear. A possible explanation is that in addition to the vertical component of the magnetic field $B_z$, the horizontal component $B_y$ also plays a role in stabilizing convection near the sidewalls. From figure~\ref{fig:Bz_By_y}(b), it can be seen that for small $\delta$ or fringe-width, although $B_y$ is strong in $y=l_y/2$ midplane, it falls sharply and becomes very small as one moves away from the midplane. Thus, for $\delta=0.01$, $B_y$ is too weak to stabilize convection near the sidewalls at $y \gg l_y/2$, resulting in sustained wall-attached convection. On the other hand, for large $\delta$, $B_y$ in the regions away from $y=l_y/2$ midplane is not as weak as that for small $\delta$. Thus, as $\delta$ is increased up to $\delta=1$, $B_y$ becomes stronger and suppresses wall-attached convection. On further increasing $\delta$, although $B_y$ grows only marginally, $B_z$ becomes too weak to suppress wall-attached convection. Hence, we get prominent wall modes again for $\delta=3$ and 9. The non-monotonic behavior of wall-modes will be explored in detail in a future study.

\section{Summary and conclusions}\label{sec:Conclusion}
In this paper, we performed detailed numerical simulations of turbulent convection of a horizontally extended domain under the influence of fringing magnetic fields. The magnetic field is generated in the gap between semi-infinite poles of permanent magnets, with the convection domain located near the edge of the gap. The quasi-static approximation of magnetohydrodynamics is applied. We kept the Rayleigh number, Prandtl number, and the maximum Hartmann number fixed at $\Ray=10^5$, $\Pran=0.021$, and $\Hazmax=120$ respectively, and varied the fringe-width by controlling the gap $\delta$ between the magnetic poles and the thermal plates. We analysed the spatial distribution of large-scale convection patterns, heat and momentum transport, and wall-attached convection, and their variations with the magnetic field profiles.

The convection patterns are found to comprise of structures whose horizontal dimensions are larger than the domain height. These structures do not exhibit any preferred orientation in regions of weak magnetic field, but become aligned perpendicular to the longitudinal sidewalls in the fringe zone. Further, the flow progressively becomes anisotropic towards the vertical direction in the fringe zone.

We obtained the best-fit relations for local Reynolds and Nusselt number as a function of the local Hartmann number that are valid for any fringe-width. We integrated these relations over the entire domain to estimate the global Reynolds and Nusselt numbers for different values of $\Hazmax$. The variation of the above global quantities with fringe-width was shown to depend on $\Hazmax$ and is governed by the balance between the convective motion in the weak magnetic flux region outside the gap between the magnets and suppression of convection in the strong magnetic flux region inside the gap between the magnets. As the fringe-width increases, the magnetic field becomes weaker in the strong magnetic flux region; it becomes stronger in the weak magnetic flux region. For large $\Hazmax$, the increase of convection in the strong magnetic field region is unable to compensate for the suppression of convection in the weak magnetic flux region, resulting in a decrease of the global heat and momentum transport with increasing fringe-width. However, for small $\Hazmax$, the increase of convection in the strong magnetic field region is more than the decrease of convection in the weak magnetic flux region, resulting in an increase of global heat and momentum transport with increasing fringe-width. 

We further analyzed wall-attached convection in the regions of strong magnetic flux. We showed that the amplitudes of the wall modes away from the corners decrease when $\delta$ is decreased from 9 to 1, but begins to increase on further reduction of $\delta$. We believe that the horizontal components of the magnetic field play a crucial role in the stability of the wall modes. A detailed study on the this behavior will be conducted in a future work.

\textcolor{black}{Here, we also remark that the initial conditions play a minor role in the evolution of the large-scale structures. In Appendix~\ref{sec:IC}, we compare therefore some of our results for $\delta=9$ with another simulation of the same setup with identical parameters, but with different initial conditions. We observe that although there is a marginal difference in the alignment of the superstructures, the other results such as the global Nusselt and Reynolds numbers, and the variations of anisotropy and the local heat transport along $y$ remain largely unchanged.}

Our present work provides important insights into the dynamics of thermal convection under spatially varying magnetic fields, which may find applications in several industrial and astrophysical flows. It should therefore be considered as a first step \textcolor{black}{towards understanding thermally-driven flows in applications where nonhomogeneous magnetic fields are encountered}. For example, our results are expected to help in designing engineering applications such as cooling blankets in fusion reactors.  Although we worked on a small set of parameters, we expect our results to hold for higher Rayleigh numbers as well. In the future, we plan to conduct a similar analysis for fluids at different Prandtl numbers.

\section*{Acknowledgements}
The authors thank A. Pandey, Y. Kolesnikov, M. Brynjell-Rahkola, \textcolor{black}{M. K. Verma, and A. Ranjan} for useful discussions. \textcolor{black}{The authors are grateful to the anonymous referees for their suggestions on improving the manuscript}. The authors gratefully acknowledge the computing time provided to them on the high-performance computer Noctua2 at the NHR Center PC$^2$ of Paderborn University (Germany). These are funded by the Federal Ministry of Education and Research and the state governments participating on the basis of the resolutions of the GWK for the national high performance computing at universities (www.nhr-verein.de/unsere-partner). \textcolor{black}{The authors further acknowledge the computating time provided by Leibniz Supercomputing Center, Garching, Germany, under the projects ``pn49ma'' and ``pn68ni''. The authors thank} the Computing Center of Technische Universit{\"a}t Ilmenau for the resources provided to them for postprocessing and visualization of their simulaion data.

\section*{Funding}
The work of S.B. is sponsored by a Postdoctoral Fellowship of the Alexander von Humboldt Foundation, Germany. 

\section*{Declaration of interests}
The authors report no conflict of interest.

\section*{Author ORCIDs}
Shashwat Bhattacharya https://orcid.org/0000-0001-7462-7680 

Thomas Boeck https://orcid.org/0000-0002-0814-7432 

Dmitry Krasnov https://orcid.org/0000-0002-8339-7749 

J{\"o}rg Schumacher https://orcid.org/0000-0002-1359-4536

\appendix
\section{\textcolor{black}{Effect of initial conditions}} \label{sec:IC}
\textcolor{black}{In this section, we explore the impact of initial conditions on the solutions in our convection setup as described in \S~\ref{sec:Model}. In the present work, all the simulations were initialized with the conduction profile for temperature and a small perturbation of amplitude $A=0.001$ in the $z$-direction for velocity. Here, we consider another simulation of our setup for $\delta=9$, $\Hazmax=120$, but with the pure convective state solution (that is, the solution for $\Hazmax=0$) as the initial condition. Like the previous runs, this simulation was initially run on a coarse grid of $480 \times 960 \times 30$ points for 100 free-fall time units in which it converged to a statistically steady state. Following this, the mesh was successively refined to the resolution of $4800 \times 9600 \times 300$ grid points and the simulation was allowed to converge after each refinement. Once the simulation reached the statistically steady state at the highest resolution, it was run for a further 21 free-fall time units and a snapshot was saved after every free-fall time unit. Henceforth, we refer to this simulation as IC~2 and the earlier simulation for $\delta=9$ (with the quiscent state being the initial condition) as IC~1.}

\begin{figure}
  \centerline{\includegraphics[scale = 0.35]{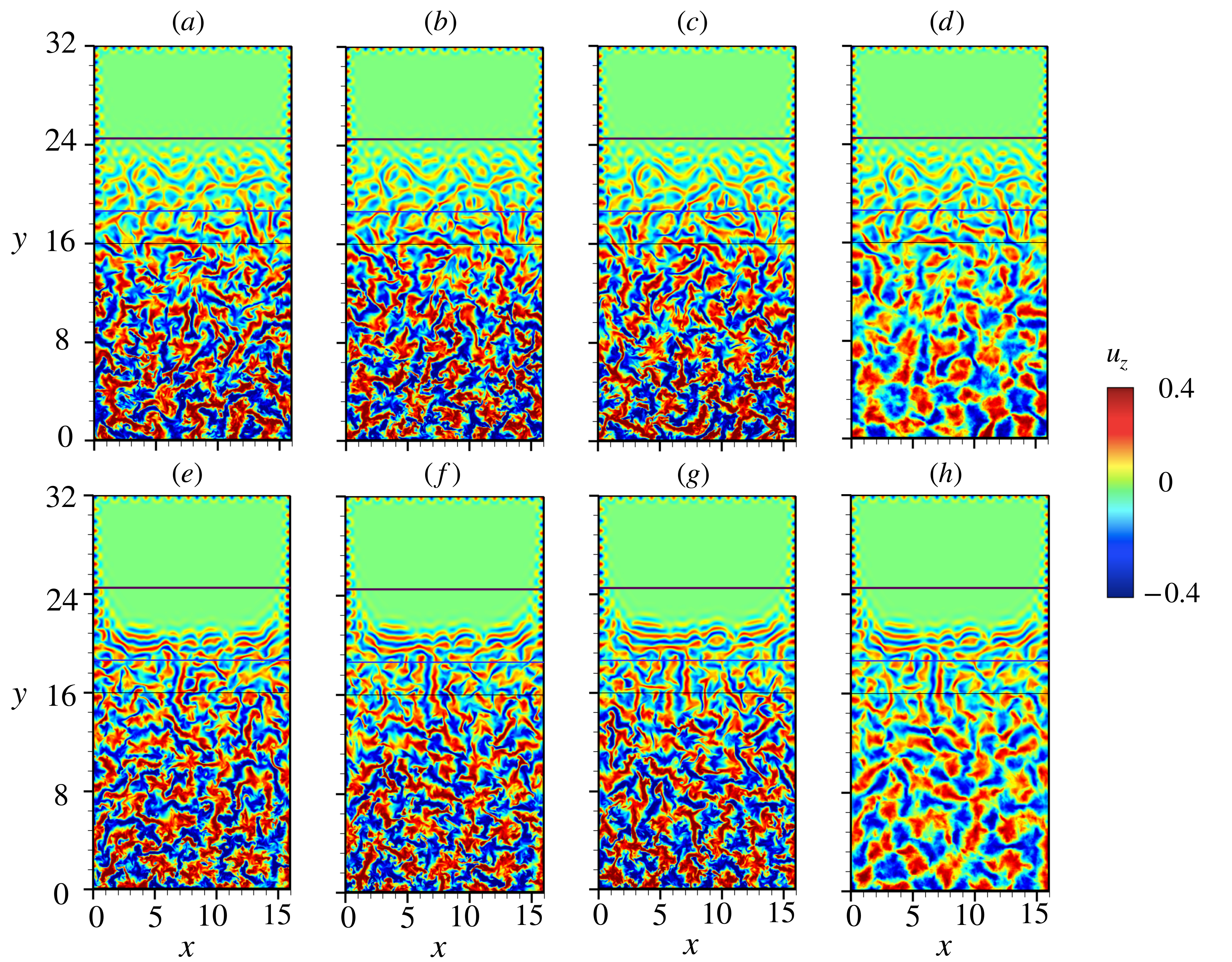}}
  \caption{\textcolor{black}{Contour plots of vertical velocity $u_z$ for $\delta=9$ in the horizontal midplane for different initial conditions. For the simulation IC~2, contour plot of $u_z$ at (a) $t=0$, (b) $t=9$, and (c) $t=18$, where $t$ is measured from the time-stamp of the first snapshot saved at statistically steady state. For the simulation IC~1, contour plot of $u_z$ at (e) $t=0$, (f) $t=9$, and (g) $t=18$, where $t$ is measured from the time-stamp of the first snapshot saved at statistically steady state. Contour plots of time-averaged vertical velocity for (d) IC~2 and (h) IC~1. The positions corresponding to $\Ha_z(y)=\Ha_{z,c}$, $\Ha_z(y)=0.8 \Ha_{z,c}$, and the edge of the magnetic poles are represented by purple, blue, and black horizontal lines, respectively, in decreasing order of thickness. The magnetic poles extend from $y=16$ to $y=\infty$.}}
\label{fig:IC_superstructures}
\end{figure}
\textcolor{black}{In figures~\ref{fig:IC_superstructures}(a)--(c), we show contour plots of vertical velocity $u_z$ on the horizontal midplane for IC~2 at three different snapshots corresponding to $t=0$, $t=9$, and $t=18$, respectively; here, $t$ is measured from the time-stamp of the first snapshot. In figure~\ref{fig:IC_superstructures}(d), we show the density plot of $u_z$ averaged over 21 free-fall time units. For comparison, we also exhibit the contour plot of vertical velocity on the horizontal midplane for IC~1. In figure~\ref{fig:IC_superstructures}(e)--(g), the plots correspond to the snapshots at $t=0$, $t=9$, and $t=18$ respectively, where $t$ is measured from the time-stamp of the first snapshot. Figure~\ref{fig:IC_superstructures}(h) exhibits the density plot of vertical velocity averaged over 20 free-fall time units. All the figures show that the large-scale patterns are qualitatively similar in the weak magnetic flux region for both IC~1 and IC~2. In the fringe zone, however, there is a difference in the alignment of the superstructures with the reorientation of the patterns perpendicular to the longitudinal sidewalls being more prominent for IC~1. Further, in the region where $\Ha_z \approx \Ha_{z,c}$, convection is completely suppressed in the bulk for IC~1 but some traces of convection are still present for IC~2. We remark that minor effects of initial conditions on magnetohydrodynamic flows have already been observed in the past; see, for example, \citet{Sukoriansky:EF1986} and \citet{Zikanov:JFM2019}.}

\begin{figure}
  \centerline{\includegraphics[scale = 0.33]{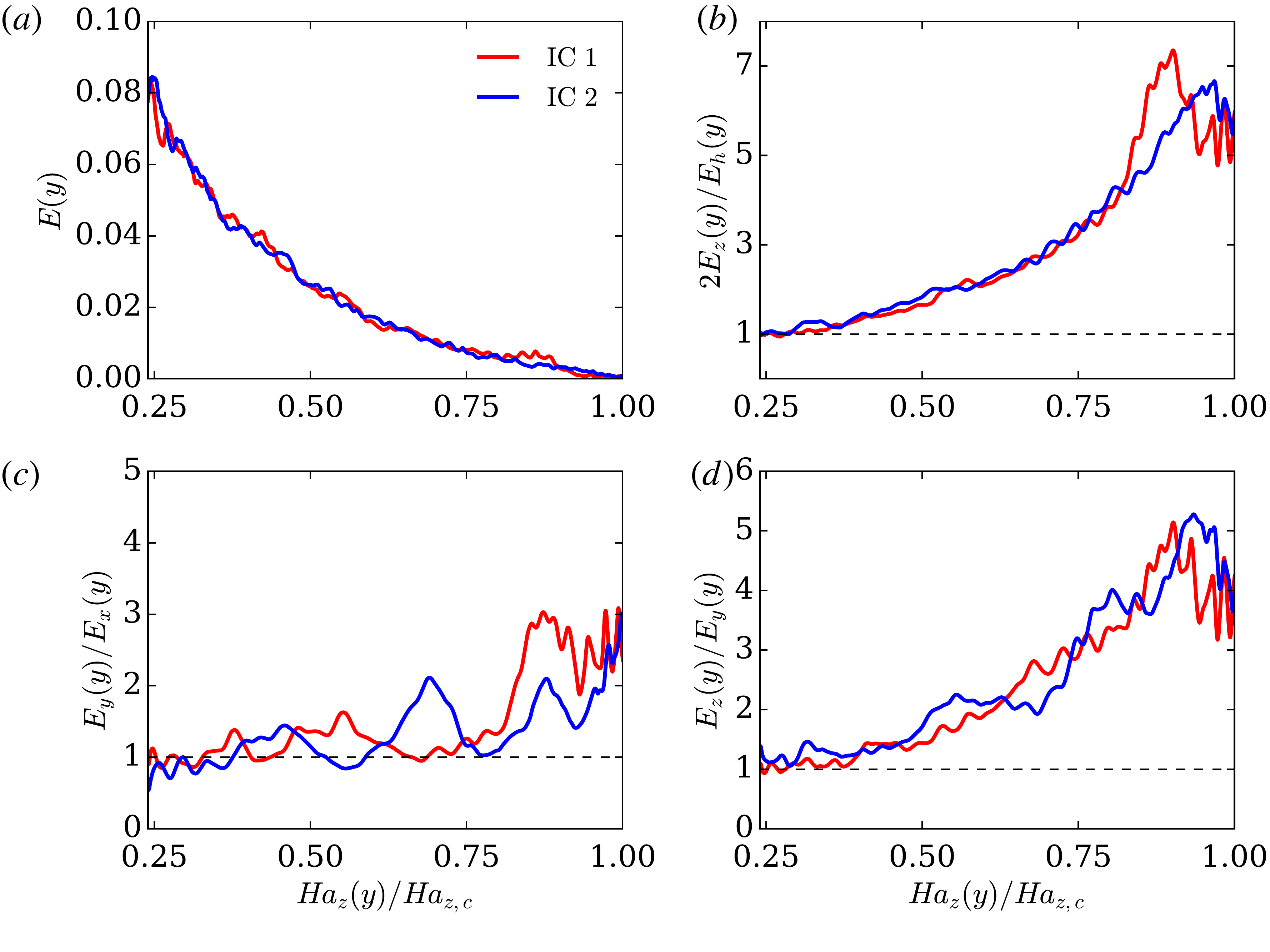}}
  \caption{\textcolor{black}{For $\delta=9$ and different initial conditions: Mean profiles of (a) the local kinetic energy $E(y)$, (b) the local vertical anisotropy parameter $2E_z(y)/E_h(y)$, (c) the local horizontal anisotropy parameter $E_y(y)/E_x(y)$, and (d) the ratio of the kinetic energy along $z$ and $y$-directions versus the Hartmann number, $\Ha_z(y)$, based on the mean vertical magnetic field $B_z(y)$. For both sets of initial conditions, the variations of $E(y)$ with $\Ha_z(y)$ are identical and the anisotropy parameters increase with the local vertical magnetic field strength.}}
\label{fig:E_Ha_IC}
\end{figure}
\textcolor{black}{We compute the planar kinetic energy ($E(y) =  0.5 \langle u_x^2 + u_y^2 + u_z^2 \rangle_{x,z}$), the local vertical anisotropy parameter ($2E_z(y)/E_h(y)$), the local horizontal anisotropy parameter ($E_y(y)/E_x(y)$), and the ratio of the kinetic energies along the $z$ and $y$ directions for IC~1 and IC~2. The aforementioned anisotropy parameters are defined in \S~\ref{sec:Superstructures}. We plot the above quantities versus the local Hartmann number in figures~\ref{fig:E_Ha_IC}(a)--(d). Figure~\ref{fig:E_Ha_IC}(a) shows that the $E(y)$-$\Ha_z(y)$ curves for IC~1 and IC~2 collapse, implying that the local variations of kinetic energy do not change with the initial conditions. From figures~\ref{fig:E_Ha_IC}(b)--(d), it is clear that although the curves of the anisotropy factors for the two different initial conditions do not collapse, they still follow the same trend. Similar to IC~1, the vertical velocity fluctuations for IC~2 become stronger compared to the horizontal ones with the increase of $\Ha_z(y)$. Further, for both IC~1 and IC~2, the velocity fluctuations in the $y$-direction become stronger compared to those in the $x$-direction as $\Ha_z(y)$ increases.}

\begin{figure}
  \centerline{\includegraphics[scale = 0.35]{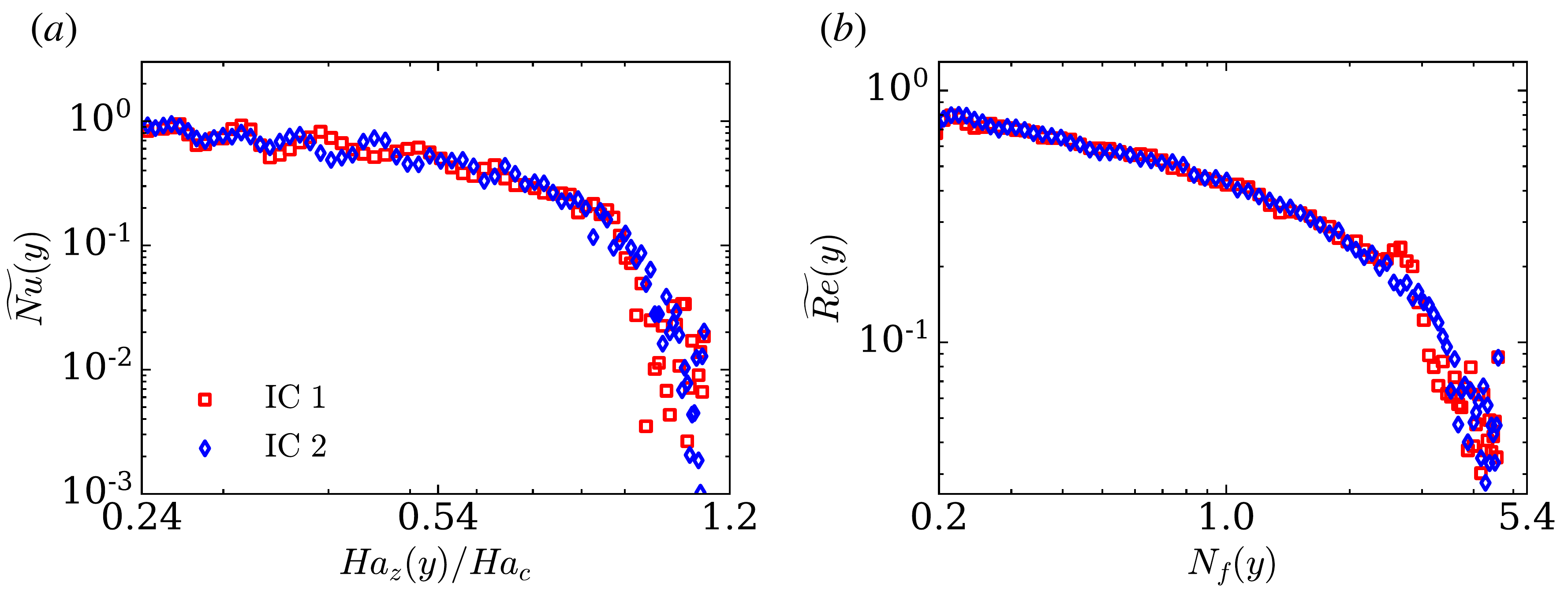}}
  \caption{\textcolor{black}{Variations of the normalized local planar Nusselt and Reynolds numbers with the \textcolor{black}{normalized} local vertical magnetic field strengths. For IC~1 and IC~2: (a) normalized local Nusselt number, $\widetilde{\Nu}(y)$, versus the normalized local Hartmann number, $\Ha(y)/Ha_c$, based on vertical magnetic field strength; and (b) normalized local Reynolds number, $\widetilde{\Rey}(y)$, versus the local free-fall interaction parameter, $N_{f}(y)$.}} 
\label{fig:Ha_ReNu_continuous_IC}
\end{figure}

\textcolor{black}{We plot the normalized local Nusselt number $\widetilde{\Nu}(y)$ versus the normalized local vertical Hartmann number $\Ha_z(y)/\Ha_{z,c}$ for both sets of initial conditions in figure~\ref{fig:Ha_ReNu_continuous_IC}(a). For the above data, we also plot the normalized Reynolds number $\widetilde{\Rey}(y)$ versus the local free-fall interaction parameter $N_f(y)$ in figure~\ref{fig:Ha_ReNu_continuous_IC}(b). The quantities $\widetilde{\Nu}(y)$, $\widetilde{\Rey}(y)$ and $N_f(y)$ are defined in equations~(\ref{eq:Nu_normalized}), (\ref{eq:Re_normalized}), and (\ref{eq:Nf}), respectively. Figures~\ref{fig:Ha_ReNu_continuous_IC}(a,b) show that both $\widetilde{\Nu}(y)$ and $\widetilde{\Rey}(y)$ for the two different initial conditions collapse well, showing that the dependence of the above quantities on the local magnetic field is independent of the initial conditions. The time-averaged global Nusselt and Reynolds numbers for IC~2 are calculated to be $\Nuglobal=2.60$ and $\Reglobal=479$. These values are nearly identical to those for IC~1 where $\Nuglobal=2.60$ and $\Reglobal=474$.}

\textcolor{black}{Our results show that the although the initial conditions can have some impact on the size and alignment of the convection patterns in the fringe zone, they do not alter the global heat and momentum transport, the local variations of heat and momentum transport, and the trend in the variations of local anisotropy.}

\bibliographystyle{jfm}

\end{document}